\documentclass[
 reprint,
 superscriptaddress,
 amsmath,amssymb,
 aps,
 pre,
]{revtex4-1}

\usepackage{hyperref} 
\usepackage[caption=false,subrefformat=parens,labelformat=parens]{subfig}  
\hypersetup{
    colorlinks = true,
    linktocpage=true,
    linkcolor = blue,
}

\usepackage{graphicx}
\usepackage{dcolumn}
\usepackage{bm}

\graphicspath{{./figures/}}

\begin{document}

\title{An exploratory study of the glassy landscape near jamming}

\author{Claudia Artiaco}
\email{cartiaco@sissa.it}
\affiliation{SISSA and INFN - via Bonomea 265, I-34136, Trieste, Italy}
\affiliation{Abdus Salam ICTP - Strada Costiera 11, I-34151, Trieste, Italy}
\author{Paolo Baldan}
\email{paolo.baldan@uniroma1.it}
\affiliation{Dipartimento di Fisica, Sapienza Universit\`a di Roma, P.le A. Moro 2, I-00185 Roma, Italy}
\author{Giorgio Parisi}
\affiliation{Dipartimento di Fisica, Sapienza Universit\`a di Roma, P.le A. Moro 2, I-00185 Roma, Italy}
\affiliation{Nanotec-CNR, UOS Rome, Sapienza Universit\`a di Roma, P.le A. Moro 2, I-00185 Roma, Italy}
\affiliation{INFN, Sezione di Roma 1,  P.le A. Moro 2, I-00185 Roma, Italy}

\begin{abstract}

We present the study of the landscape structure of hard and soft spheres as a function of the
packing fraction and of the energy. We find that, on approaching the jamming transition, the number of
different configurations available to the system has a steep increase and that a hierarchical organization
of the landscape emerges. We use the knowledge of the structure of the landscape to predict the values
of thermodynamic observables on the edge of the transition.
\end{abstract}

\maketitle


\section{\label{sec:introduction}introduction}

\noindent In this exploratory study, we investigate the properties of the landscape near jamming starting from independent configurations in the same cage. We follow the line of research of \cite{charbonneau2015jamming, charbonneau2015numerical,berthier2016growing}. We look at the landscape local minima both at jamming and in the overcompressed region. The overcompressed phase near the jamming point is expected to present a similar physics to hard spheres at jamming \cite{charbonneau2012universal,franz2017universality, berthier2011microscopic}: in particular, we expect to find the same values for most of the critical exponents. Therefore, in both cases, we expect to find a Gardner-like landscape \cite{kurchan2013exact, charbonneau2014exact, charbonneau2014fractal, charbonneau2017glass, biroli2018liu, scalliet2019marginally}, in which each cage is broken into a fractal hierarchy of subcages.  Beyond the Gardner transition, the replica symmetry becomes continuously broken (full-RSB) \cite{mezard1987spin} and the subcages are organized in an ultrametric structure.

The presence of the Gardner transition in finite-dimensional glasses is still a matter of debate: renormalization group studies yield controversial results \cite{moore2011disappearance,urbani2015gardner,charbonneau2017nontrivial}, while numerical simulations suggest that the existence of the Gardner transition may be model dependent \cite{degiuli2014effects,berthier2016growing,scalliet2017absence,seoane2018spin,liao2019hierarchical,scalliet2019nature}. However, in hard spheres systems the jamming transition presents universal critical properties which seems to be independent both on the preparation protocol and on the dimension of the system \footnote{Anyway, in finite dimensions new effects might come into play, due, for instance, to the \textit{rattlers} and the \textit{bucklers} \cite{charbonneau2015jamming}.}, showing the same features from $d=\infty$ to $d=2$ \cite{charbonneau2017glass}.

We present a direct inspection of the existence of the Gardner phase in spheres systems in three dimensions near jamming, looking at the properties of the landscape local minima. We study the packing fraction and the energy distribution of the local minima packings, and their relative distance in the configuration space, both at jamming and in the overcompressed region. We look for the ultrametric structure of the landscape at jamming. Furthermore, we predict how the shape of key thermodynamical observables modifies when approaching the transition point both in temperature and in pressure.

Our analysis shows that the system undergoes a roughness transition, which brings about isostaticity on approaching jamming \cite{charbonneau2014fractal}. The transition is characterized by a steep increase in the number of local minima that, at the jamming point, are organized in an ultrametric way. The deepest minima are close both in terms of the packing fraction and of the distance in the configuration space. They also have large basins of attraction. Moreover, in the overcompressed region, the cumulative of the number of local minima at small energies behaves as a power-law with a packing fraction dependent exponent.

The plan of the paper is as follows. In Sec.~\ref{sec-general-framework}, we describe the general framework and the model. The results of the numerical simulations on the landscape structure at jamming are presented in Sec.~\ref{sec-jamming-landscape}. In Sec.~\ref{sec-overcompressed}, we describe the results in the overcompressed region. Sec.~\ref{sec-properties-finite-T-P} is devoted to the extrapolation of the landscape properties at finite temperatures and pressures and, in  Sec.~\ref{sec-conclusions}, we summarize our findings and we discuss the perspectives future studies.

\section{\label{sec-general-framework} General Framework}

\noindent This Section is devoted to summarizing the main features of the hard spheres phase diagram in infinite dimensions \cite{parisi2010mean,charbonneau2017glass}.

The control parameters of a hard spheres system are the pressure and the packing fraction  $\varphi$ \footnote{The packing fraction $\varphi$ is defined as the fraction of the volume occupied by the particles. In a monodisperse system, $\varphi=\frac{4}{3}\pi r^3\rho$, where $\rho=\frac{N}{V}$ is the density and $r$ is the radius of the spheres.}. Different regions of the phase diagram are identified by the behavior of the mean-square-displacement (MSD) ${\Delta(t)= \frac{1}{N}\sum_{i=1}^N|\textbf{r}_i(0)-\textbf{r}_i(t)|^2}$. 
When the pressure is small, the system is ergodic and $\underset{t\to\infty}{\lim} \Delta(t)=\infty$. An equilibrium compression of the liquid can be carried out only up to $\varphi_d$, the dynamical transition point. At $\varphi_d$, the phase space becomes clustered into an exponential number of glassy states. These clusters are mutually inaccessible and trap the dynamics at infinite times: $\underset{t\to\infty}{\lim} \Delta(t)=\Delta_{\text{liq}}<\infty$. For $\varphi_g>\varphi_d$, it is possible to define a restricted equilibrium state \cite{franz1995recipes}, known as the stable glass phase: the system can completely relax inside the metastable state but long-time diffusion is forbidden. The particles of a stable glass are caged by their neighbors and vibrate around their ``amorphous equilibrium positions'' in \textit{cages} whose typical size is $\Delta_{\text{liq}}$.

Compressing further the stable glass up to $\varphi_G(\varphi_g)$, one encounters the Gardner transition where even the restricted equilibrium is lost: the stable glass state breaks into a hierarchical structure of marginal states (landscape marginal stability, LMS). This implies the existence of delocalized soft modes, diverging susceptibilities \cite{biroli2016breakdown} and anomalous rheological properties \cite{yoshino2014shear}.

Finally, compressing the system up to the point where the pressure diverges, the system jams. At jamming, the packings are mechanically rigid and isostatic, meaning that the number of mechanical constraints is equal to the number of degrees of freedom \cite{o2003jamming}. Isostaticity implies that the system is mechanically marginally stable (MMS). Hence, at the jamming point, the number of soft modes is enhanced.

Validating the mean-field picture for finite-dimensional systems would greatly increase the global understanding of the glass transition \cite{debenedetti2001supercooled,cavagna2009supercooled,berthier2011theoretical}. In finite dimensions, the dynamical transition reduces to a crossover because the energy barriers between metastable states are finite. For this reason, it is possible to numerically generate equilibrated glassy configurations even at $\varphi_g>\varphi_d$ via improved Monte Carlo simulations, known as Swap Monte Carlo \cite{gazzillo1989equation,grigera2001fast,ninarello2017models}. Nevertheless, in conventional dynamics, the time spent by the system in a metastable state can be, to a good extent, considered infinite. 

\subsection{\label{subsec:methods} Methods}
\begin{figure}[htp]
\centering
\includegraphics[width=0.485 \textwidth]{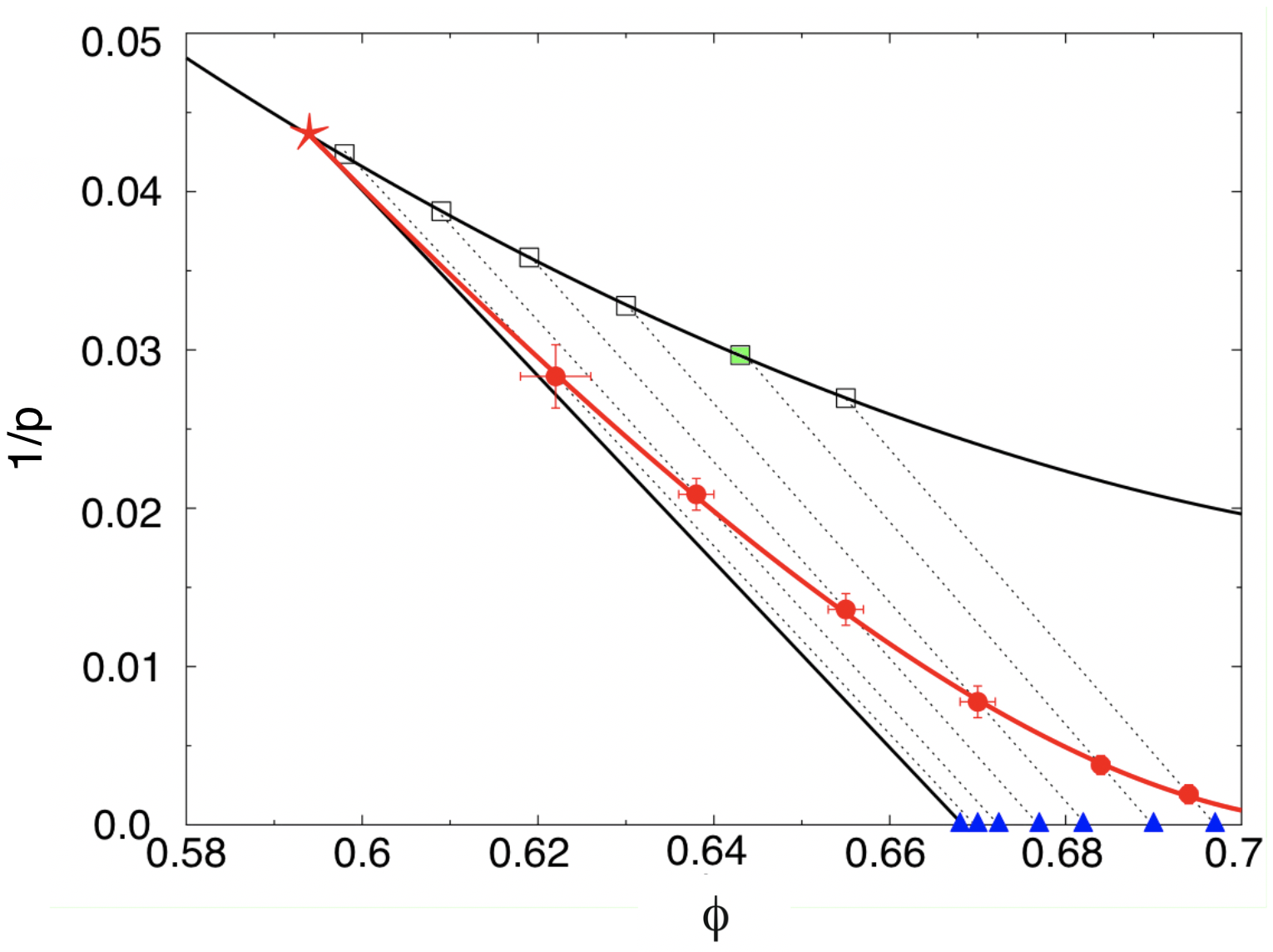}
\caption{(Color online) Phase diagram of a polydisperse hard spheres system in $3d$. The initial glassy configurations are equilibrated in correspondence of the green square. The green and the white squares lie on the equation of state (EOS) of the liquid at various $\varphi_g$; the red line represents the Gardner transition line, $\varphi_G(\varphi_g)$, and the line with blue triangles represents the jamming line, $\varphi_J(\varphi_g)$. Adapted from \cite{berthier2016growing}.}
\label{fig-phase-diagram}
\end{figure}

\noindent Beyond the Gardner transition the number of minima of each cage is expected to diverge in a way that depends exponentially on the number of components of the system \cite{stillinger1999exponential}. Therefore, we choose to restrict our study to 100 3$d$ spheres. In order to enhance the equilibration process, the spheres diameters are drawn from the continuous probability distribution  $P(\sigma)\propto\sigma^{-3}$ with $\sigma_{min}/\sigma_{max}=0.45$ \cite{ninarello2017models}. The equilibration has been achieved via the constant-pressure Swap Monte Carlo algorithm \footnote{The equilibration beyond the dynamical transition point via the Swap Monte Carlo dynamics has been carried out by the Montpellier research group and L. Berthier, that we warmly thank.} to produce 5 glassy configurations at ${\varphi_g=0.647}$ ($\varphi_d=0.594$), which are in different cages and whose position in the phase diagram is represented by the green square of Fig.~\ref{fig-phase-diagram}. We study the landscape near jamming starting from independent glassy configurations in the same cage.

The absence of spatial order and long-time diffusion are key features of glasses. Hence, we firstly study the structure of the 5 initial glassy configurations, computing the radial distribution function \cite{hansen1990theory}, and we investigate their dynamical properties, measuring the MSD as a function of time. Our analysis does not show any sign of crystalline order, phase separation, and long-time diffusion.

\subsubsection{Samples Generation and Compression Protocols}

\noindent Using an NVT Monte Carlo dynamics \footnote{In a sweep of the NVT Monte Carlo dynamics, we propose the displacement of each particle of the system. If the displaced particle overlaps another particle, the movement is refused. We measure the time in units of Monte Carlo sweep.}, we evolve in time each of the starting configurations. At $t=\tau_{cage}$, the system enters the caging regime, signaled by a plateau in the MSD. Saving the time-evolved configurations every $3\cdot\tau_{cage}$ sweeps, we create a set of independent configurations belonging to the same cage, called \textit{clones}. The set of \textit{clones} generated from the same initial configuration is called \textit{sample}. To prevent long-time diffusion and the breaking of the cage, we periodically restart the dynamics from the starting configuration. Indeed, due to the stochasticity of the Monte Carlo dynamics, we are guaranteed to sample different trajectories at each restart. We end up with a set of $2\times10^5$ independent glassy configurations in the same cage.

Furthermore, via a fast compression \footnote{We move the particles via the NVT Monte Carlo dynamics and, after a fixed number of steps, we increase all the diameters of a factor $\gamma=10^{-3}$. If in the new packing some spheres do overlap, we further move the particles. The procedure stops when the packing fraction of the configuration reaches $\varphi=0.68$. We call this as \textit{fast} compression because it carries the system out-of-equilibrium.} of the starting configuration, we generate a new glassy configuration at higher packing fraction, $\varphi=0.68$. The fast compression does not distort the jamming landscape, but pushes the system into one of the subcages, if, at that packing fraction value, the cage is already broken into subcages. From the new configuration, we produce another set of $2\times10^5$ packings in the same (sub)cage. Therefore, studying the jamming landscape from a sufficiently compressed configuration, we expect to retrieve a subset of the local minima found starting from the equilibrated glassy configuration \footnote{This method can also be employed to study the heights of the landscape barriers. Given a $\varphi>\varphi_g$, one can generate many compressed configurations at $\varphi$ via independent fast compressions and, from each of the compressed configurations, generate a sample. Each sample can then be brought to the jamming transition via LP. Repeating the procedure at increasing values of $\varphi$ and studying which landscape local minima survive in different samples, a full map of the landscape can be constructed, including the heights of landscape barriers.}. 

Thus, we follow three compression protocols: (i) taking each sample at $\varphi=0.647$, we bring each clone to the jamming point, via an instantaneous  LP compression (App.~\ref{app-lp}); (ii) we repeat the same procedure for each sample at $\varphi=0.68$; (iii) we take $10^5$ clones from each sample at $\varphi=0.647$ and we bring them at several packing fraction values beyond the jamming point. Then, we locally relax them via the FIRE algorithm \cite{bitzek2006structural}.

\subsubsection{Studying the Landscape at Jamming}

\noindent To study the landscape at jamming, we reach the jamming point via the Linear Programming (LP) algorithm \cite{donev2004linear, torquato2010robust} (App.~\ref{app-lp}). LP works in the uncompressed region \footnote{The uncompressed region is known as the SAT phase in the context of constraint satisfaction problems, while the overcompressed region is called the UNSAT phase \cite{franz2017universality}.}, meaning that overlaps among particles are not allowed during the compression protocol. 

In the uncompressed region, the probability of having hard spheres packing at $\varphi$ is $\propto e^{-\frac{Np}{\varphi}}$, where $p$ is a proxy for the pressure and $N$ is the system size. Hence, the jamming transition is reached at $p=\infty$ and the jammed packings maximize $\varphi$.

In this exploratory study, we present the analyses of the jamming landscape of 4 cages, reaching the jamming point both from $\varphi=0.647$ and $\varphi=0.68$.

\subsubsection{Studying the Landscape in the Overcompressed Region}

\noindent To investigate the overcompressed region it is necessary to introduce a soft sphere potential. We choose to employ a harmonic repulsive potential \cite{o2003jamming}
\begin{equation}
\label{eq-harmonic-potential}
    U(\{r\})=\frac{\epsilon}{2}\sum_{i<j}\Bigl(1-\frac{r_{ij}}{\sigma_{ij}}\Bigl)^2\theta{(\sigma_{ij}-r_{ij})} \, ,
\end{equation}
where $r_{ij}=| \boldsymbol{r}_{i}-\boldsymbol{r}_j|$ is the distance between the centers of two particles, $\sigma_{ij}=\frac{\sigma_i+\sigma_j}{2}$ is the sum of their radii and $\epsilon$ is a constant that fixes the energy unit. 

We can study how the system behaves in the overcompressed phase when a finite temperature $\beta^{-1}$ is introduced, weighting each energy minimum with its Gibbs measure $\propto e^{-\beta E}$. Temperature is measured in units of $\epsilon/k_B$.

\section{\label{sec-jamming-landscape} Landscape at Jamming}
\noindent In this Section, we present our results on the study of the landscape local minima at jamming. Notice that larger values of the jamming packing fraction $\varphi_J$ correspond to minima which are \textit{deeper} in the landscape; while smaller values of $\varphi_J$ coincide with minima \textit{higher} in the landscape.

Most of the jammed packings obtained using the LP algorithm are isostatic (App.~\ref{app-rattler}), \textit{i.e.} they verify $N_c=N_c^{\text{iso}}\equiv(N-1)d+1$ \footnote{ $N_c=(N-1)d+1$ is the isostaticity condition for a finite system under periodic boundary conditions (PBC) \cite{charbonneau2015jamming}. Here, $N_c$ is the number of contacts of the jammed packing, $N$ is the number of particles (excluding the rattlers (App.~\ref{app-lp})) and $d$ is the dimension of the system.}, consistently with the mean field solution. Since several works \cite{lerner2013low,wyart2012marginal,goodrich2012finite,hopkins2013disordered,charbonneau2015jamming} have highlighted the importance of having $N_c=N_c^{\text{iso}}$ to observe key aspects of jamming criticality, in this study we restrict our analysis to the packings with $N_c=N_c^{\text{iso}}$.

\subsection{Local Minima}
\label{sec-local-minima}

\noindent Compressing the clones of a sample up to jamming, we find many jammed packings with the same $\varphi_J$. This is especially true for high values of $\varphi_J$. Furthermore, the packings with the same $\varphi_J$ present also the same arrangement of the particles but for a rigid translation \footnote{The rigid translation is due to the translational invariance of the system. The effect has been taken into account in all the results shown in the present study. In particular, in computing the distance $\Delta$ (Eq.~\ref{eq-square-distance}) and the overlap Q (Eq.~\ref{eq-overlap}), we subtracted the displacement of the center of mass among the two configurations.}, \textit{i.e.} they represent the same local minimum at jamming.

In each different cage, we find that the jamming local minima coming from $\varphi=0.68$ are a subset of those coming from $\varphi=0.647$, meaning that at $\varphi=0.68$ the starting cage is already broken into subcages. 

\begin{figure}[htp]
\centering
\subfloat[]
{\includegraphics[width=\columnwidth]{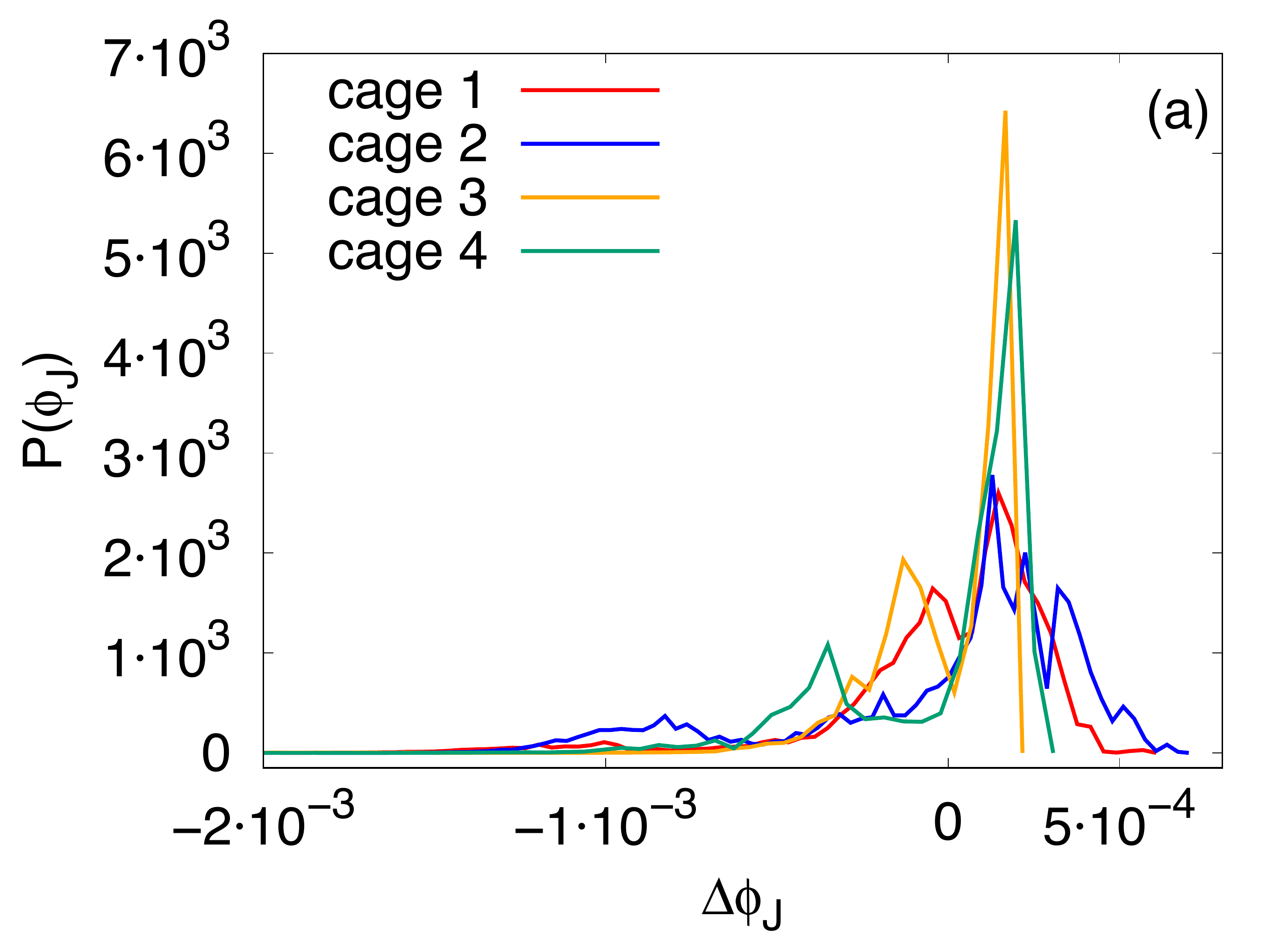}}
\vspace{-6mm}
\subfloat[]
{\includegraphics[width=\columnwidth]{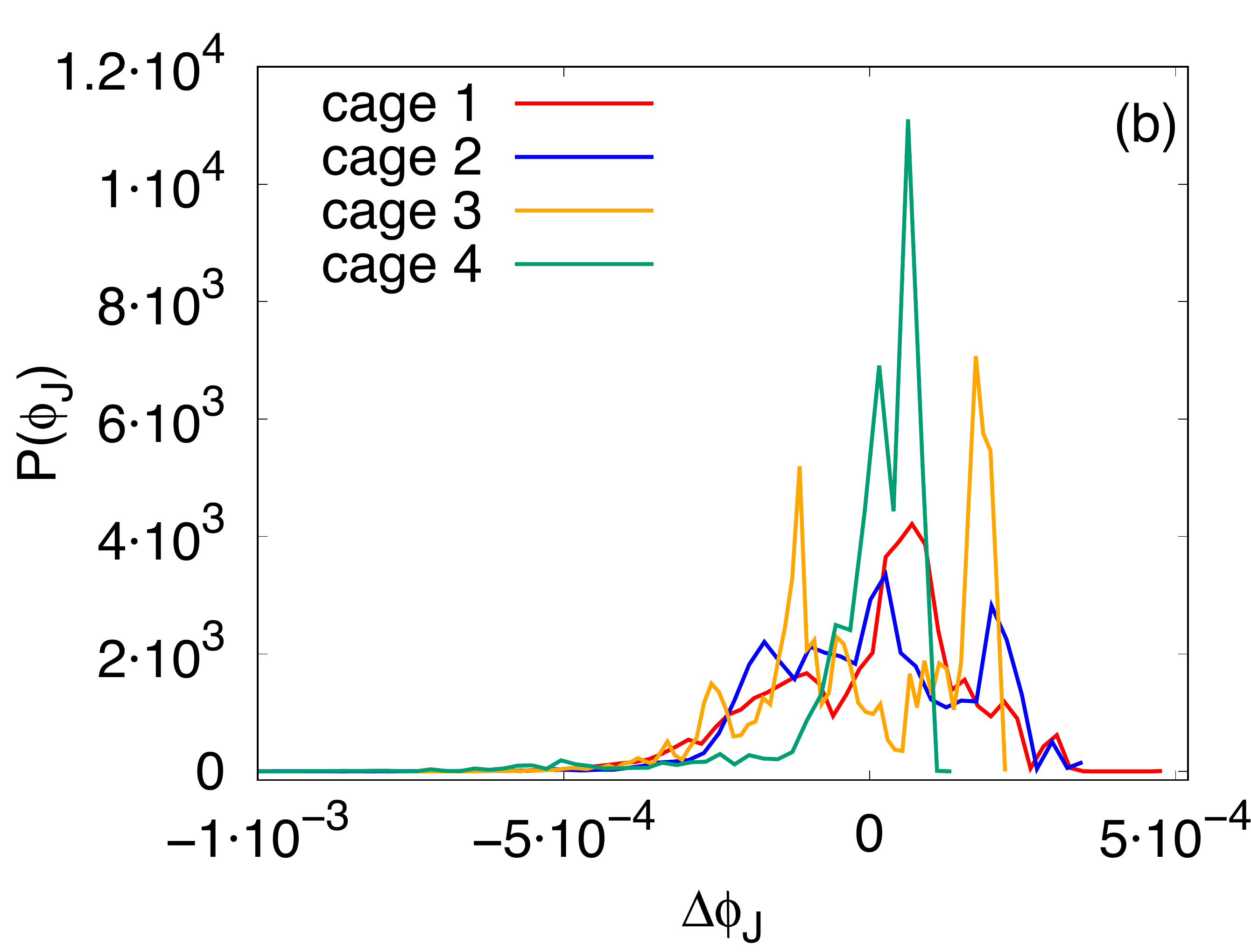}}
\caption{(Color online) Probability distribution of the local minima at jamming, obtained by compressing about $2\times10^5$ clones. On the $x$ axis, $\Delta\varphi_J=\varphi_J-\langle\varphi_J\rangle$, where $\langle\varphi_J\rangle$ is the mean jamming packing fraction of the cage considered. On the $y$ axis, the probability of finding $\varphi_{J}$. (a) Results of all the cages generating the clones at $\varphi=0.647$. (b) Results of all the cages generating the clones at $\varphi=0.68$.}
\label{fig-histo-allConf}
\end{figure}

For each sample, we compute the local minima distribution at jamming. The local minima distributions, shown in Fig.~\ref{fig-histo-allConf}, have different supports, depending on the depth of the basin of that cage. The $\varphi_J$'s distributions are not self-averaging quantities. The average $\varphi_J$ values, merging the data from different cages, are $\overline{\varphi}_{J,\,0.647}=0.6955$ and $\overline{\varphi}_{J,\,0.68}=0.6957$.  

Two important features of the landscape are summarized by  Fig.~\ref{fig-histo-allConf}: many deep local minima are found with high probability and are characterized by similar $\varphi_J$ values. We argue that the first feature means that their basins of attraction are large. 

We find that the $\varphi_J$ distributions may be wider or narrower. The average differences between the highest and the lowest jamming packing fractions are $\overline{\Delta\varphi}_{J,\,0.647}^{\,\text{max-min}}=0.0035$ and $\overline{\Delta\varphi}_{J,\,0.68}^{\,\text{max-min}}=0.0016$. The average ratios between the number of distinct local minima found and the total number of clones in the same sample are, respectively, $f_{0.647}=0.32$ and $f_{0.68}=0.20$.

Ordering in a progressive way the jamming packing fractions, we can define
\begin{gather}
    \delta_n=\varphi_{n+1}-\varphi_n\geq0\\
    0\leq r_n=\frac{\text{min}\{\delta_n,\delta_{n-1}\}}{\text{max}\{\delta_n,\delta_{n-1}\}}\leq1
\end{gather}
If the $\varphi_J$'s are uniformly distributed, one expects the level statistics to be Poissonian \cite{oganesyan2007localization} and, thus, $P_{\text{Poisson}}(r)=2/(1+r)^2$ with $\langle r \rangle_{\,\text{Poisson}}\simeq 0.386$.

Computing $\langle r \rangle$ in each sample, the Poissonian statistics turns out to be almost valid. Indeed, the average on all the samples is $\langle r\rangle_{0.647}=0.384\pm0.001$ and $\langle r\rangle_{0.68}=0.382\pm0.002$.

\subsection{Structure in the Configuration Space}
\label{subsec-jamming-structure}

\noindent Since at jamming, in each sample, we find a huge number of distinct local minima ($\simeq10^4-10^5$), here we study in detail the arrangement in the landscape only of a subset of them, the 1000 deepest minima.

\begin{figure}[htp]
\centering
\subfloat[]
{\includegraphics[width=\columnwidth]{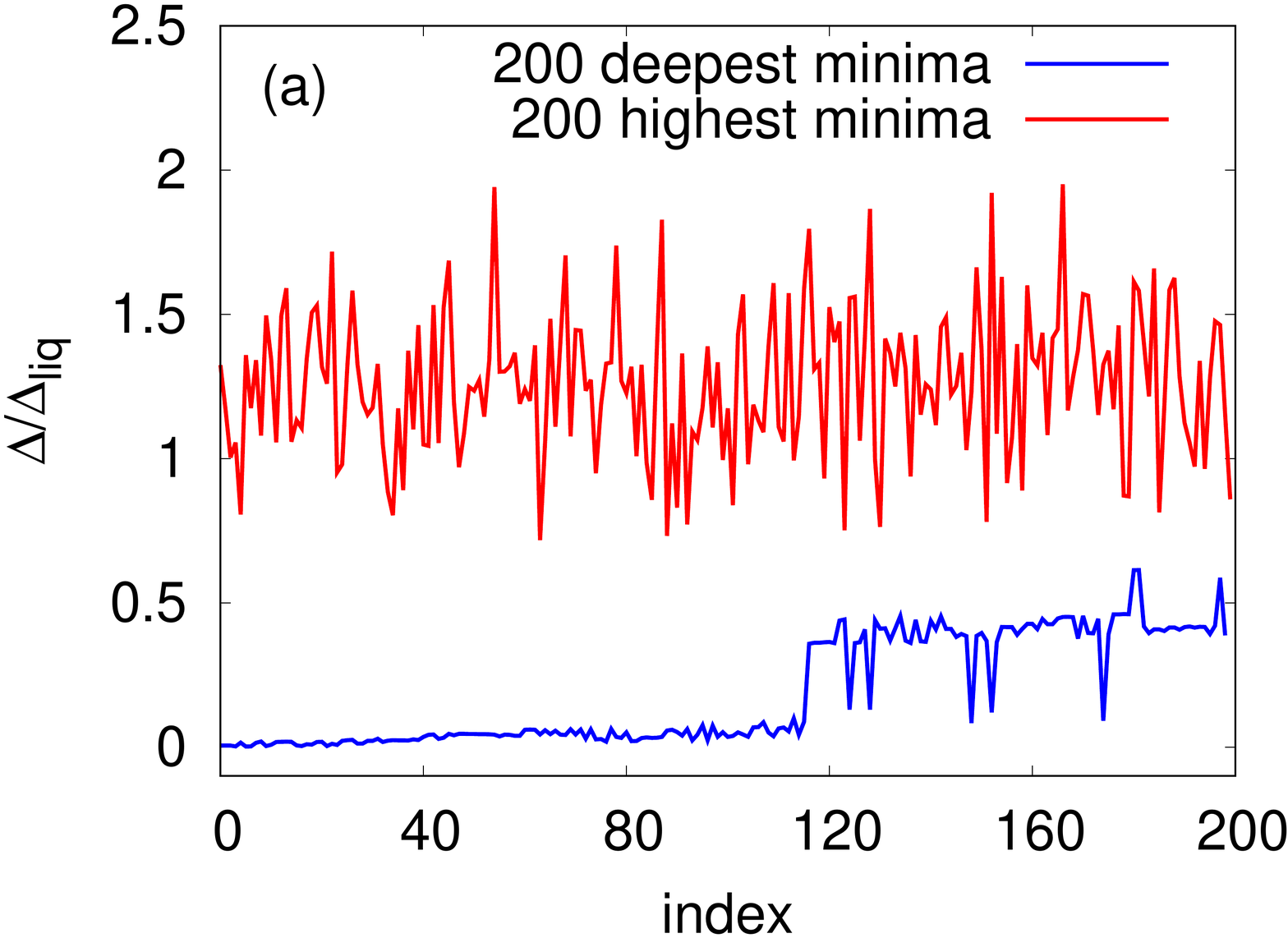}
\label{fig-delta-conf1-647-deep}}
\vspace{-8mm}
\subfloat[]
{\includegraphics[width=\columnwidth]{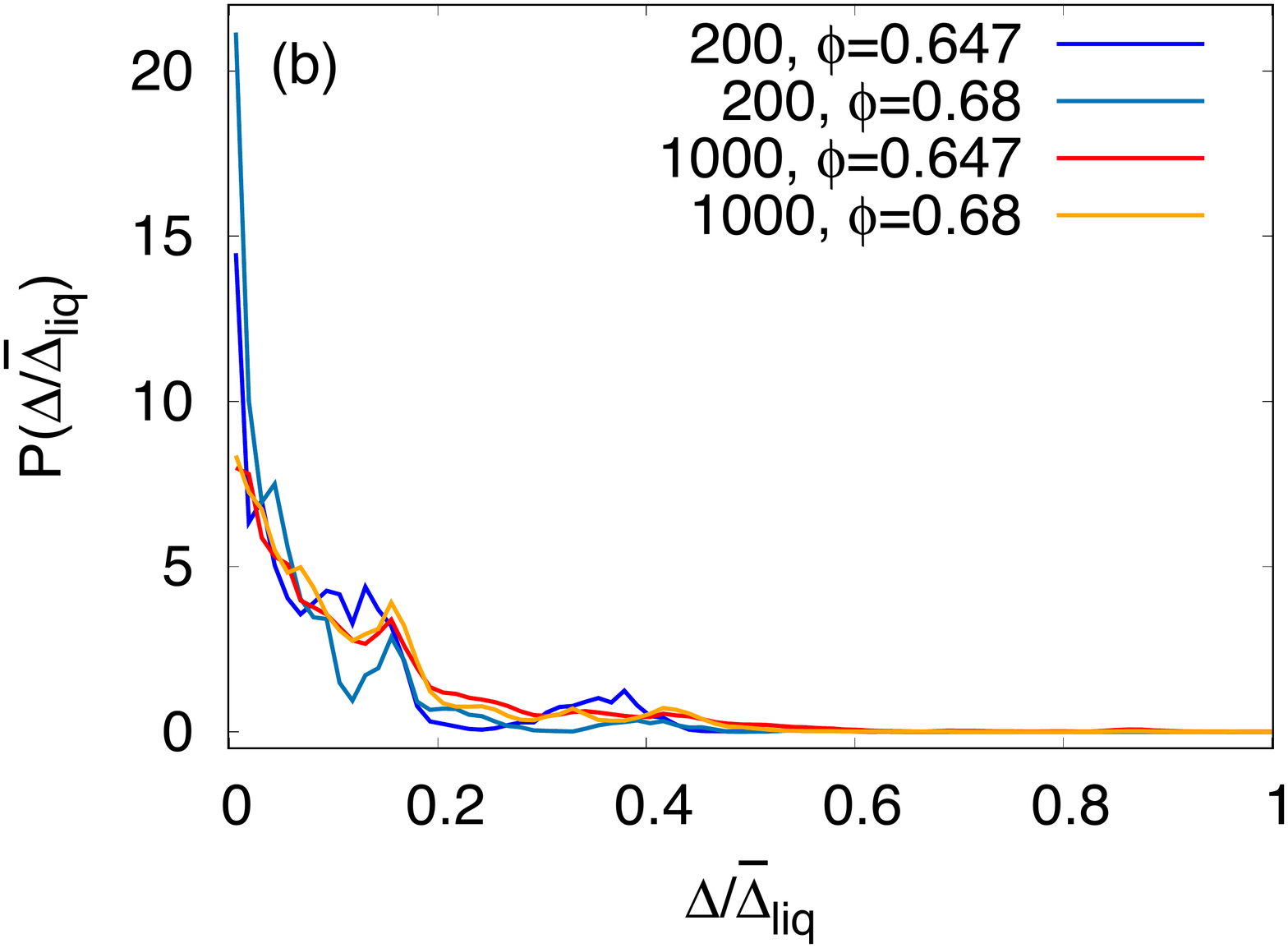}
\label{fig-delta-1000-200-reciproche}}
\caption{(Color online) (a) Data of cage 1 generating the clones at $\varphi=0.647$. On the $y$ axis, the distances $\Delta$ among the deepest minimum and the 200 deepest and 200 highest minima of the same sample, normalized to the MSD of cage 1 ($\Delta_{\text{liq}}=0.0066$). On the $x$ axis, the index increases as $\varphi_J$ decreases. (b) $\Delta$ distributions joining the data of all the cages. On the $x$ axis, $\Delta$ is normalized to $\overline{\Delta_{\text{liq}}}=7.3\times10^{-3}$. In blue and azure, the distance distributions among the 200 deepest minima starting the compressions up to jamming from, respectively, $\varphi=0.647$ and $\varphi=0.68$. In red and orange, the same plots considering the 1000 deepest minima of each sample.}
\label{fig-delta-deepest}
\end{figure}

First of all, we want to understand how the deepest minimum (\textit{i.e.}, the one with the highest $\varphi_J$) of each sample is located in the configuration space. To do so, we compute its square distance
\begin{equation}
    \label{eq-square-distance}
    \Delta\equiv\Delta^2_{\alpha\beta}=\frac{1}{M}\sum_{i=1}^{M}(\boldsymbol{x}_i^{\alpha}-\boldsymbol{x}_i^{\beta}-\boldsymbol{\delta})^2
\end{equation}
from the other 200 deepest and the 200 highest minima of the same sample. Here, $\alpha$ and $\beta$ are the minima index and $i$ is the particle index; $M$ is the total number of particles excluding the rattlers of both $\alpha$ and $\beta$ and $\boldsymbol{\delta}$ is the distance of the centers of mass of the two minima (Apps.~\ref{app-rattler},~\ref{app-distances}). The $\Delta$ values have to be compared with the average values of the MSD in the caging regime, which are $\overline{\Delta_{\text{liq}}}\equiv\overline{\text{MSD}_{\text{plateau}}^{0.647}}=7.3\times10^{-3}$ and $\overline{\Delta_{\text{liq}}^{0.68}}\equiv\overline{\text{MSD}_{\text{plateau}}^{0.68}}=2.6\times10^{-3}$. In Fig.~\subref*{fig-delta-conf1-647-deep}, is shown the result for cage 1 generating the clones at $\varphi=0.647$. The deepest minimum of this sample appears to be located in a deep well where there are many other very deep minima of the cage (blue line). Moreover, the deepest minimum is faraway from the highest minima (red line). We find almost the same behavior in all the samples.

In Fig.~\subref*{fig-delta-1000-200-reciproche} are shown the $\Delta$ distributions among the 200 and 1000 deepest minima, having joined the results of all the cages. It is clear that the deepest minima of the landscape in each sample are usually close one to the other.

In a sample, the configuration correspondent to the deepest minimum is usually found many times; but, it is not the one with the greatest number of occurrences. We denote the latter as \textit{the most probable minimum}.

Performing the same analyses of Fig.~\ref{fig-delta-deepest} on the most probable minimum, we find that it is close to many other highly probable minima; hence, in the landscape at jamming, there is a large basin of attraction made up by large subbasins of attraction.

Furthermore, studying in the same way the properties of the highest minima, we find that they are far apart from all the other minima and do not form a well. They have small basins of attraction. We can imagine them as narrow ponds on the walls of the landscape basins.

Another observable useful to understand the relative position of two local minima at jamming is the overlap \cite{ninarello2017models}. It is defined as
\begin{equation}
    Q\equiv Q_{\alpha\beta}=\frac{1}{M}\sum_{i,j}^{1,M}\Theta\biggl(a-\mid\boldsymbol{x}_i^{\alpha}-\boldsymbol{x}_j^{\beta}-\boldsymbol{\delta}\mid\biggr)
    \label{eq-overlap}
\end{equation}
As in Eq.~\ref{eq-square-distance}, the sum is restricted to the $M$ particles which are non-rattlers in none of the two configurations $\alpha$ and $\beta$; $\boldsymbol{\delta}$ is the distance of the centers of mass of $\alpha$ and $\beta$ (Apps.~\ref{app-rattler},~\ref{app-distances}). $\Theta$ is the Heaviside step function. We choose $a=0.03$. $Q\rightarrow1$ when the number of displaced particles decreases. 

\begin{figure}[htp]
\centering
\includegraphics[width=\columnwidth]{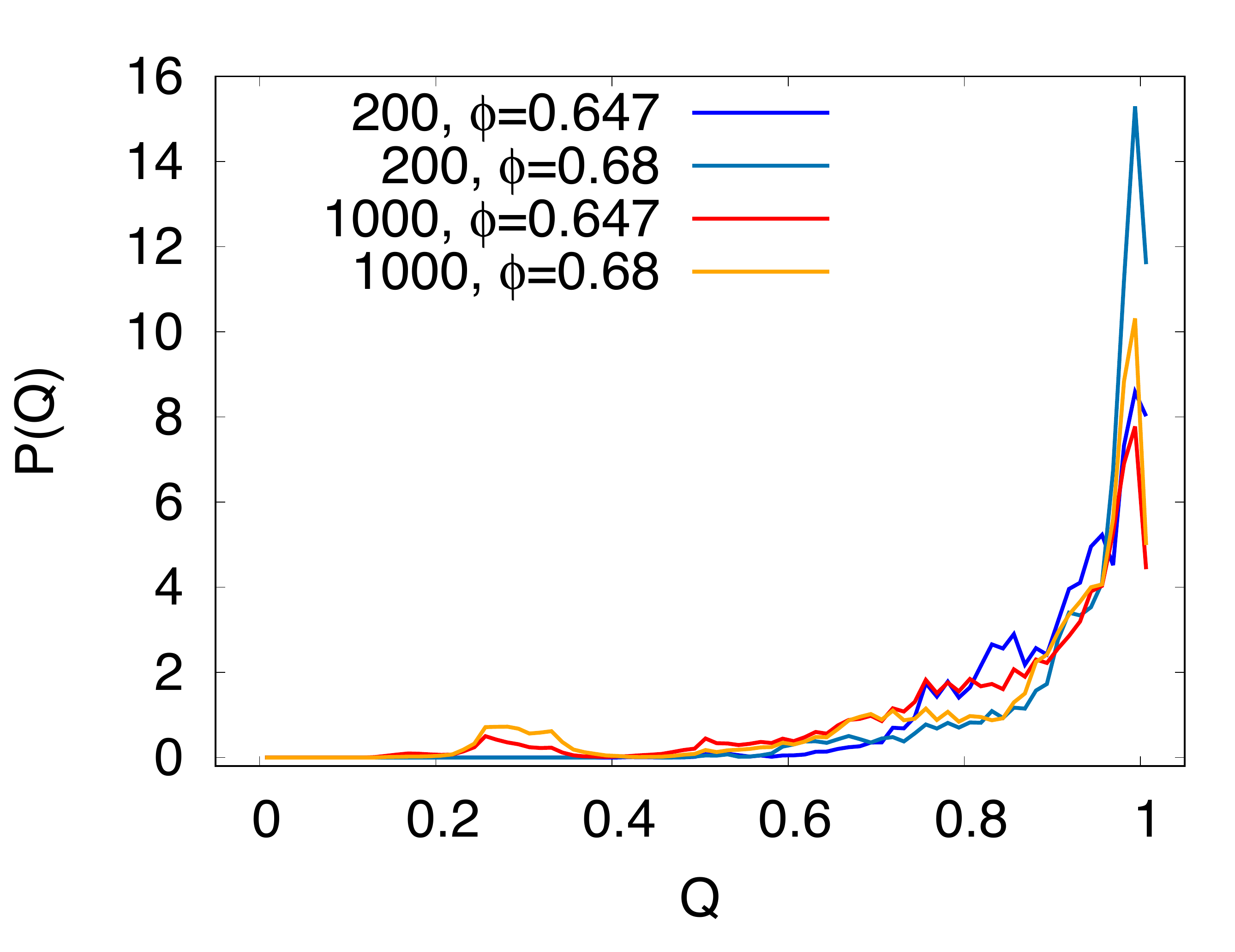}
\caption{(Color online) Probability distributions of $Q$ joining the data of all the cages. In blue and azure, the Q distributions among the 200 deepest minima starting the compressions up to jamming from, respectively, $\varphi=0.647$ and $\varphi=0.68$. In red and orange, the same plots considering the 1000 deepest minima of each sample.}
\label{fig-QAB-1000-200-reciproche}
\end{figure}

$Q$ gives complementary information to $\Delta$. For instance, a high value of $\Delta$ can be due to the presence of only one particle which has very different positions in the two packings $\alpha$ and $\beta$; if it is so, $Q$ has a large value. Otherwise, if many particles are displaced by a small amount, the same $\Delta$ can correspond to a small value of $Q$. In Fig.~\ref{fig-QAB-1000-200-reciproche} is shown the behavior of the overlap $Q$ for the same minima used in Fig.~\subref*{fig-delta-1000-200-reciproche}. In our results, given a couple of minima $\alpha$ and $\beta$ there is not a direct correspondence between a high value of $Q$ and a small value of $\Delta$. However, Fig.~\subref*{fig-delta-1000-200-reciproche} and Fig.~\ref{fig-QAB-1000-200-reciproche} show that the deepest minima of each sample on average are close in terms of both $Q$ (few particles are displaced) and $\Delta$ (small distance). This confirms our description of the landscape structure in terms of deep wells in which the deepest minima are contained. 

\subsection{Hierarchical Structure}
\label{subsec-hierarchical-structure}

\noindent According to the mean-field picture, we expect to find an ultrametric structure of the basins. To verify this hypothesis, we construct the heatmaps \footnote{The heatmaps are made with R \cite{R}, using the \textit{average method}.} of the 1000 deepest minima of each sample, using the distance $\Delta$ as the dissimilarity measure. 

\begin{figure}[htp]
\centering
\subfloat[]
{\includegraphics[scale=0.3]{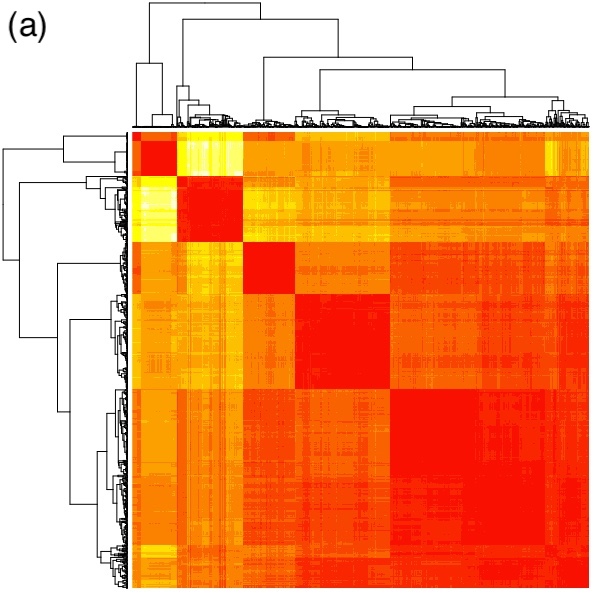}}
\hspace{5mm}
\subfloat[]
{\includegraphics[scale=0.3]{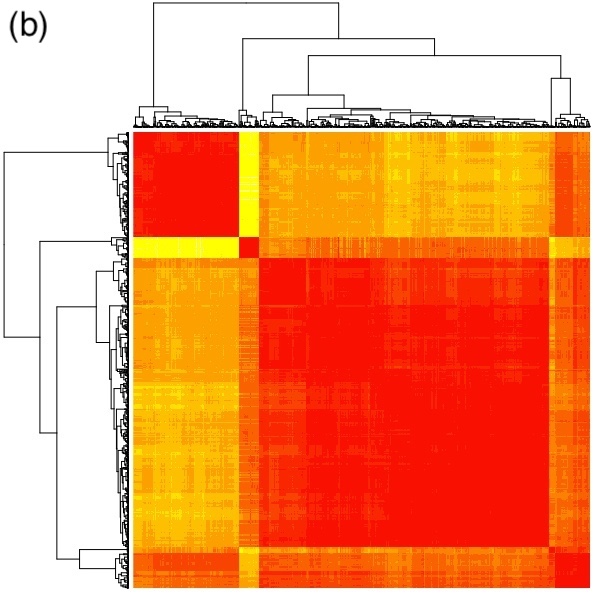}}
\caption{(Color online) Heatmaps constructed with the 1000 deepest minima of a sample. (a) Heatmap for cage 1 coming from $\varphi=0.647$. $\text{AC}=0.996$. (b) Heatmap for cage 3 coming from $\varphi=0.68$. $\text{AC}=0.997$. The clusters are created using the average method.}
\label{fig-heatmaps}
\end{figure}

In Fig.~\ref{fig-heatmaps} are shown the heatmaps of cage 1 and cage 3 for the samples at $\varphi=0.647$ and $\varphi=0.68$, respectively. To provide a quantitative characterization of the clustering properties of the selected minima, we compute the agglomerative coefficient (AC) for each heatmap  \footnote{{$\langle \text{AC}= 1-m(i) \rangle$}. For each observation $i$, $m(i)$ is the dissimilarity to the first cluster it is merged with divided by the dissimilarity of the merger in the final step of the algorithm.}. The average on all the samples is $\overline{\text{AC}_{0.647}}=0.994$ and $\overline{\text{AC}_{0.68}}=0.995$. These high values of AC tell us that the minima have a good clustering structure and, so, present a nearly ultrametric structure.

\subsection{Interim Discussion: The Main Features of the Landscape at Jamming}

\noindent Summarizing, the main features of the landscape at jamming are:

\begin{itemize}
    \item All the samples show a huge number of distinct local minima at jamming. The landscape appears very complex.
    
    \item The deepest minima of the landscape are found more often than the highest, and so we argue that their basins of attraction are larger.
    
    \item The dendrograms constructed with the 1000 deepest minima of each sample have good clustering properties (high values of AC). This is evidence in favor of an ultrametric structure of the landscape. In particular, the landscape seems to be more ultrametric when we reach the jamming point coming from $\varphi=0.68$. Notice that a rigorous ultrametric structure can be seen only considering all the minima of a basin. Indeed, when we reach jamming from $\varphi=0.68$, we have pushed the initial equilibrated configuration into a subcage: we are sampling a smaller basin and we expect to detect better the ultrametricity. 
    
    \item The arrangement of the local minima is compatible with several studies on disordered systems \cite{parisi2003statistical}: the deepest minima are close in the configuration space and form a large deep well; instead, the highest minima do not form a well and have small probabilities of being found.
    
    \item A small square distance $\Delta$ often means a high $Q$ value: there are few displaced particles between the two configurations. Anyway, there are cases in which a lot of particles are slightly moved.
    
    \item The level statistics at jamming seems to be determined by Poisson statistics, meaning that there is no level-repulsion.
    
\end{itemize}

In the next Section, we will study the landscape in the overcompressed region where, near the jamming point, we expect to find a similar physics to hard spheres at jamming \cite{charbonneau2012universal,franz2017universality, berthier2011microscopic}.

\section{Landscape in the Overcompressed Region}
\label{sec-overcompressed}

\noindent Since the point at which a sample jams depends on the preparation protocol \cite{chaudhuri2010jamming}, we independently determine the jamming packing fraction $\varphi_J$ for each cage through compression-decompression cycles \cite{charbonneau2015jamming}, extrapolating the point at which the energy reaches zero. We then instantaneously inflate each clone to an excess packing fraction $\varphi_e = \varphi - \varphi_J$, ranging from $\varphi_e=5 \times 10^{-3} \, \varphi_J$ to $\varphi_e=100 \times 10^{-3}  \, \varphi_J$, and we locally relax it to the closest energy minimum, \textit{i.e.}, the inherent structure \cite{stillinger1982hidden}, via a  FIRE routine \cite{bitzek2006structural}.

\subsection{Local Minima}
\label{sec-IS-properties}

\begin{figure}[h]
    \centering
    \includegraphics[width=\columnwidth]{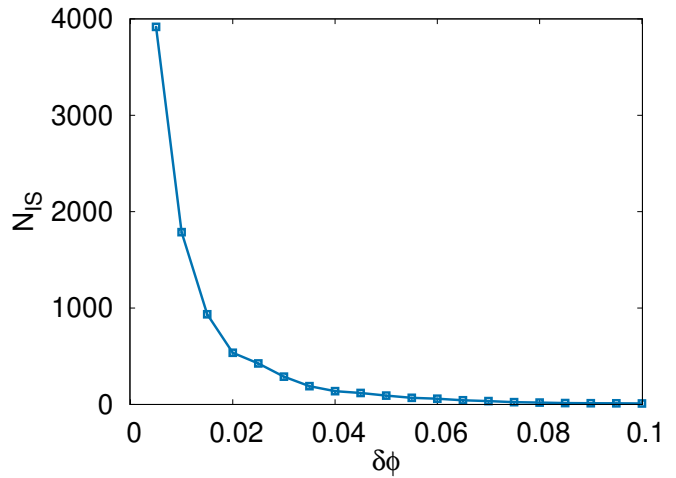}
    \caption{(Color online) Number of inherent structures as a function of the excess packing fraction $\delta\varphi$. In performing the computation, we coalesced minima whose difference in energy was less than $10^{-14}$. On approaching the jamming point, the number of different minima increases abruptly and seems to diverge at jamming.}
    \label{fig:nstates}
\end{figure}

\noindent In Fig.~\ref{fig:nstates} we depict the number of different inherent structures, $N_{\text{IS}}$, as a function of $\delta\varphi=\frac{\varphi_e}{\varphi_J}$. Fig.~\ref{fig:nstates} clearly shows that, as jamming is approached, the number of distinct inherent structures steeply increases, passing from a few units at the highest packing fractions to several thousand, consistently with a roughening of the landscape at jamming.

\begin{figure}[h]
    \raggedleft
    \includegraphics[width=\columnwidth]{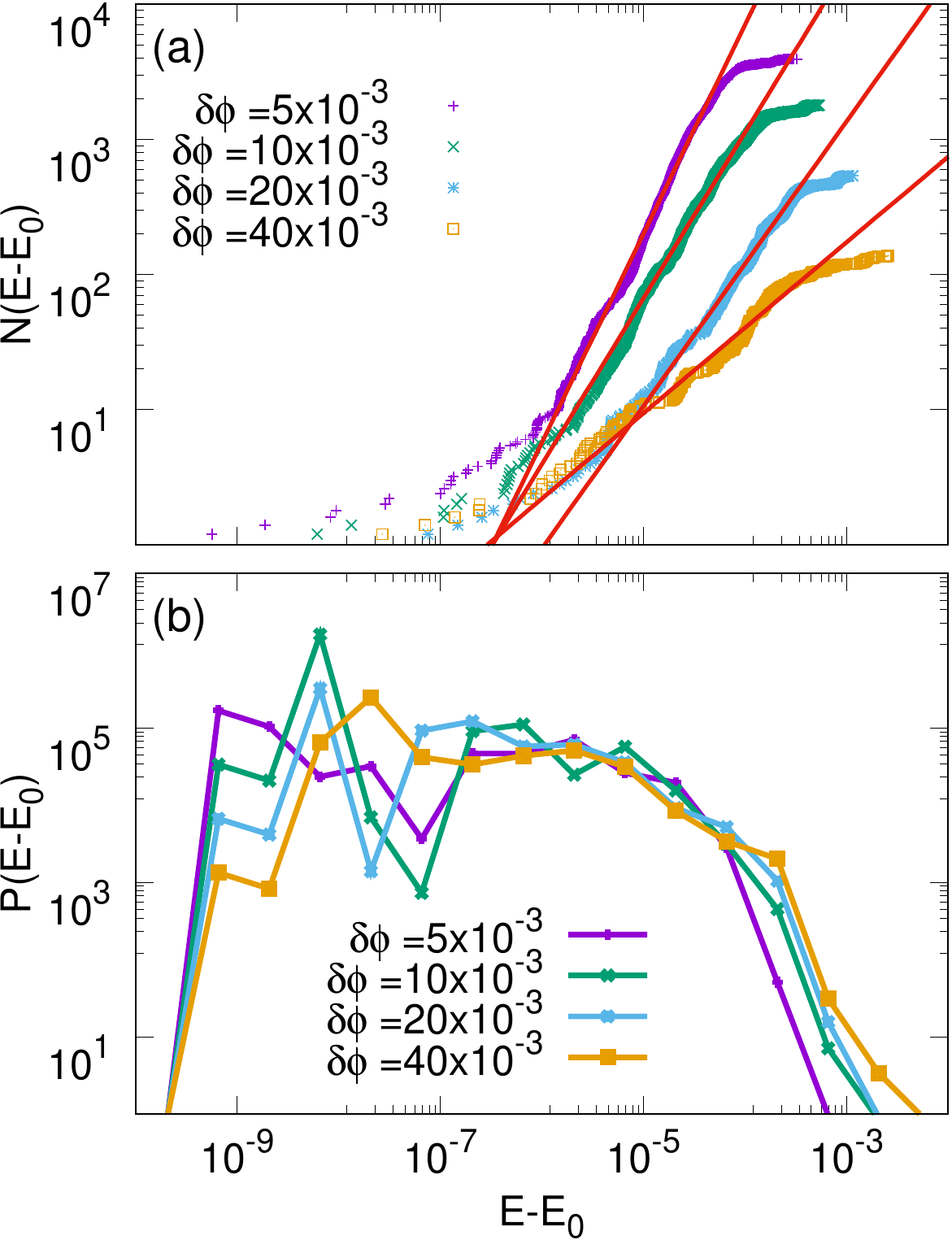}
    \caption{(Color online) (a) Cumulative distribution of the number of inherent structures, averaged over different cages, as a function of the energy difference $E-E_0$, where $E_0$ is the energy of the deepest minimum of the cage. The red lines corresponds to power-law fits with exponents $\alpha = 1.54,\, 1.22,\, 1.05,\, 0.63$. (b) Probability distribution of the energy of the inherent structures for the same packing fraction values of (a). The energy minima that lie in the region where the cumulative saturates are $10^2$ times less likely than those belonging to the power-law regime.}
    \label{fig:complexity}
\end{figure}

\begin{figure}[h]
    \includegraphics[width=\columnwidth]{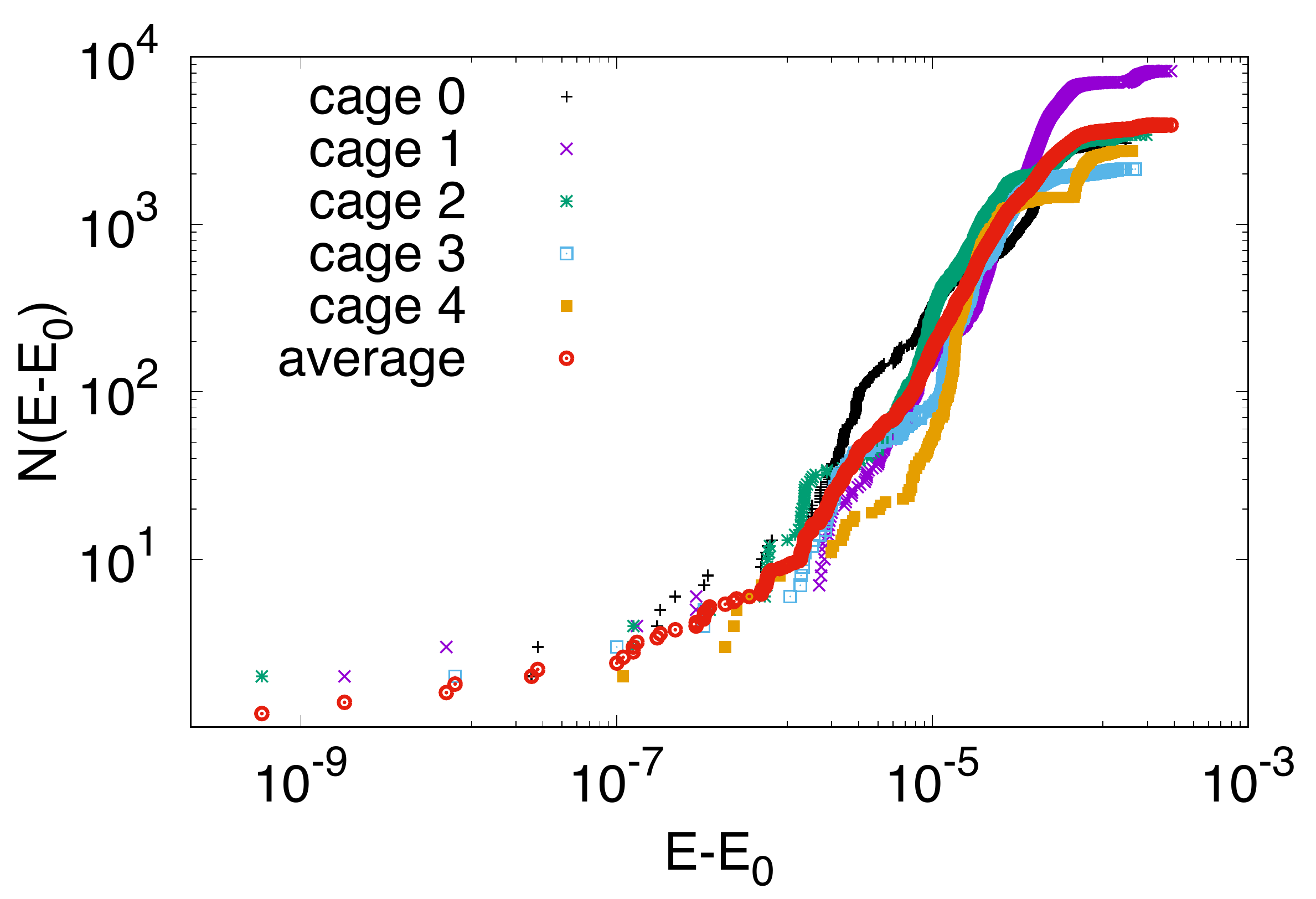}
    \caption{(Color Online) Cumulative of the number of inherent structures as a function of $E-E_0$, where $E_0$ is the energy of the deepest minimum of the cage, for the 5 cages at $\delta \varphi = 5\times 10^{-3}$.}
    \label{fig:complexity_fluct}
\end{figure}

In Fig.~\hyperref[fig:complexity]{\ref{fig:complexity}(a)} and Fig.~\ref{fig:complexity_fluct}, we show the cumulative distributions of the number of inherent structures as a function of the inherent structure energy \footnote{Note that $\log{N(E)}$ is related to the configurational entropy of the system.}, $N(E-E_0)$ ($E_0$ denotes the energy of the deepest minimum of the cage). Fig.~\hyperref[fig:complexity]{\ref{fig:complexity}(a)} presents the results averaged over all the samples for several packing fraction values, ranging from $\delta \varphi = 5 \times 10^{-3}$ to $\delta \varphi = 40 \times 10^{-3}$, while in Fig.~\ref{fig:complexity_fluct} we show the behavior in different samples at $\delta\varphi=5\times10^{-3}$. Three regimes are present: (i) seemingly exponential at small energies, probably in connection with finite-size effects \cite{kallus2016scaling}, (ii) power-law with a packing fraction dependent exponent $\alpha(\varphi)$ at intermediate energies, (iii) plateau at high energies.

To better understand the origin of the plateau, we look at the probability distribution of the inherent structure energies, in
Fig.~\hyperref[fig:complexity]{\ref{fig:complexity}(b)}. It appears highly concentrated around values close to the deepest minimum of the landscape. It follows that the highest energy minima have very narrow basins of attraction \cite{sciortino2005potential, martiniani2016turning} that are visited with very small probability by the equilibrium dynamics, even though they contribute significantly to the configurational entropy of the system \cite{parisi2003statistical}. Therefore, in Figs.~\hyperref[fig:complexity]{\ref{fig:complexity}(a)}-\ref{fig:complexity_fluct} the saturation of $N(E-E_0)$ at high energies is likely due to the inability of the algorithm to sample all the minima in the cage. As a consequence, Figs.~\hyperref[fig:complexity]{\ref{fig:complexity}(a)}-\ref{fig:complexity_fluct} provide a good representation of $N(E-E_0)$ only at low and intermediate energies.

Hence, Fig.~\ref{fig-histo-allConf} and Fig.~\ref{fig:complexity} represent landscapes with the same features: the deepest minima of the cage are more likely to be found, while the highest have narrow basins of attraction and are hardly sampled by gradient-descent-like algorithms.

Moreover, Fig.~\ref{fig:complexity_fluct} gives evidence of the presence of the Gardner physics: even though the cumulative distributions $N(E-E_0)$ of the 5 starting cages have the same qualitative behavior, the total number of detected states has very large sample-to-sample fluctuations. This leads to the loss of the self-averaging property when the jamming transition is approached.

\subsection{Harmonic Properties of the Local Minima}

\noindent The analysis of the harmonic properties of the minima is carried out by studying the Hessian matrix: 
\begin{align}
\label{eq:hessian}
&H_{ij}^{\alpha\beta} = \frac{\partial^2 U}{\partial r_i^\alpha \partial r_j^\beta} =
\nonumber
\\ &= \delta_{ij} \sum_{k \in \partial i} \left[ n_{ik}^\alpha n_{ik}^\beta + \frac{\varepsilon_{ik}}{r_{ik}}\left( n_{ik}^\alpha n_{ik}^\beta -
\delta^{\alpha\beta}\right)\right] + \\
&- \delta_{\langle ij\rangle}\left[ n_{ij}^\alpha n_{ij}^\beta + \frac{\varepsilon_{ij}}{r_{ij}}\left( n_{ij}^\alpha n_{ij}^\beta - \delta^{\alpha\beta}\right)\right], \nonumber
\end{align}
with $\alpha, \beta = x,y,z$ are position vector components, $\varepsilon_{ij} = \sigma_{ij} - r_{ij}$ is the overlap between two spheres, $\vec{n}_{ij} = (\vec{r}_j - \vec{r}_i)/r_{ij}$ is a unit vector, both $\delta_{ij}$ and $\delta^{\alpha\beta}$ are Kronecker deltas, $\delta_{\langle ij\rangle}$ indicates a contact between a pair of particles, and $\partial i$ denotes the set of neighbors of $i$. 

Matrix \ref{eq:hessian} has 3 zero modes corresponding to the global translational invariance of the system plus a set of clone-dependent zero modes connected to the presence of rattlers, which are particles that are not part of the force network \cite{charbonneau2015jamming}, and therefore play no role in the thermodynamics of the system \footnote{The rattlers are removed before computing the Hessian.}. 

Given the eigenvalues of the Hessian $\{\lambda\}$, we define the logarithm of the curvature as: 
\begin{equation}
    \Gamma = \frac{1}{N_{dof}}\sum_{i=1}^{N_{dof}} \log(\lambda_i) \, , 
    \label{eq:curvature}
\end{equation}

\noindent where $N_{dof}=3(N-1-n_{ratt})$ is the total number of degrees of freedom, $n_{ratt}$ being the number of rattlers.

\begin{figure}[h]
    \centering
    \includegraphics[width=\columnwidth]{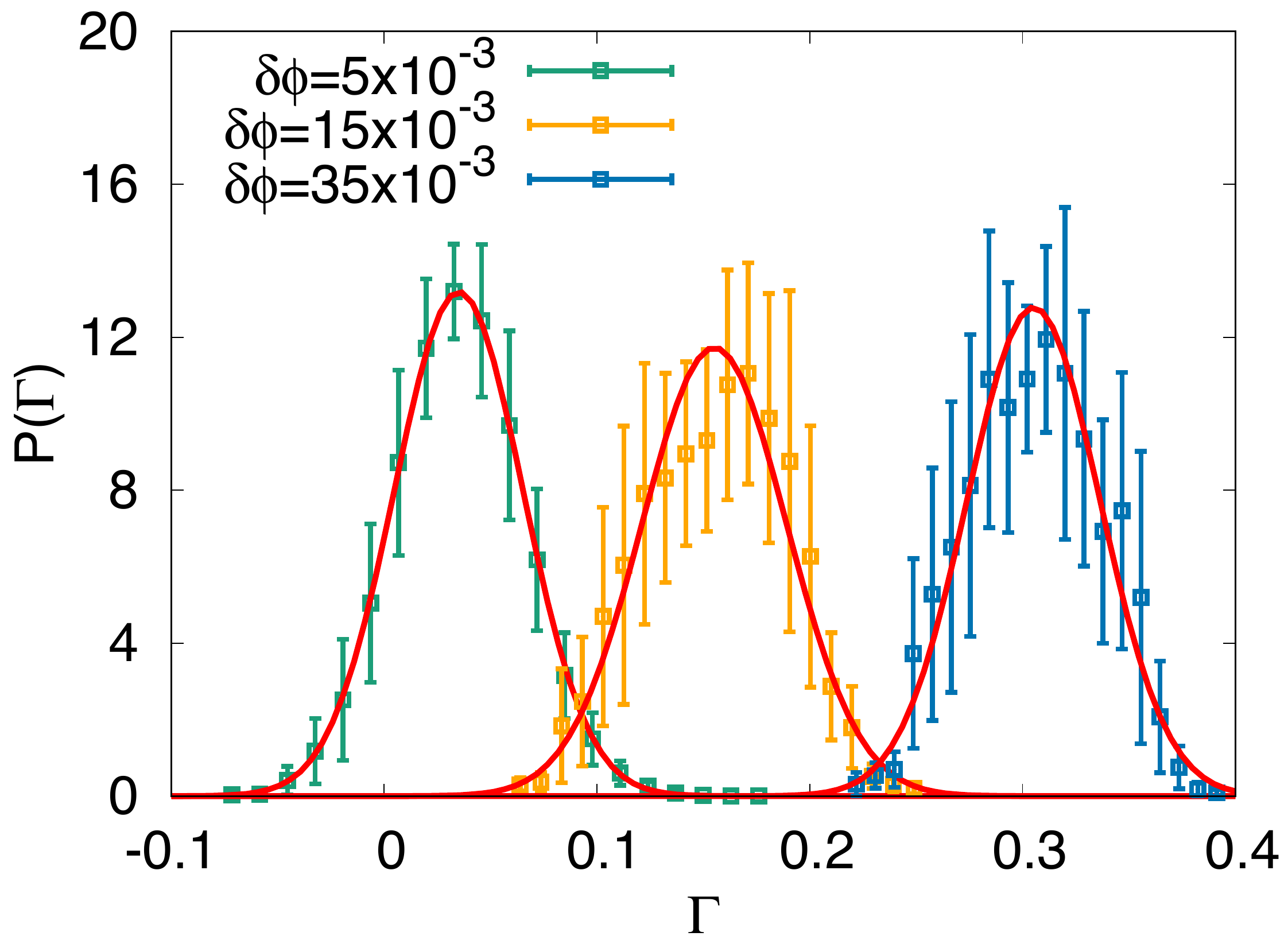}
    \caption{(Color online) Probability distribution of the curvature at ${\delta \varphi = 5,\,15,\,35 \times 10^{-3}}$. The red curves are Gaussians with the same mean and standard deviation of the experimental data. Error bars are computed with the jackknife method.}
    \label{fig:curvature}
\end{figure}

Fig. \ref{fig:curvature} tracks the evolution of the curvature as a function of the packing from $\delta \varphi= 35 \times 10^{-3}$, where a Gaussian profile first appears, to $\delta \varphi = 5 \times 10^{-3}$. At larger packing fractions, the Gaussian profile is not present. 

Gaussianity implies that large finite-size effects $O(1/\sqrt{N})$ take place approaching the jamming transition. We can also notice that the distribution shifts towards smaller values of $\Gamma$ as the packing fraction is decreased, compatibly with the onset of mechanical marginal stability at jamming. Indeed, smaller values of $\Gamma$ are associated with larger basins of attraction of the inherent structures. Error bars are computed with the jackknife method \cite{amit2005field} on the 5 samples and their height is a consequence of the large sample-to-sample fluctuations, as already highlighted in Sec.~\ref{sec-IS-properties}.

\section{\label{sec-properties-finite-T-P} Thermodynamical Properties of the Landscape}

\noindent Building on the knowledge of the landscape structure, we try to predict how physical observables would change on approaching the jamming transition at finite temperature in the overcompressed region and at finite pressure in the undercompressed one. In the first case, we start our analysis from the potential energy landscape of Sec.~\ref{sec-overcompressed}, while in the second case we start from the landscape at jamming studied in Sec.~\ref{sec-jamming-landscape}.

For two systems with the same disorder (in this context, two clones) we define the probability distribution of their distance (see Eq.~\ref{eq-square-distance} and App.~\ref{app-distances}) as
\begin{align}
    \label{eq:pdelta}
    P(\Delta) = \frac{1}{Z^2} \int Dr^{(1)}Dr^{(2)}e^{-\beta H(r^{(1)})} e^{-\beta H(r^{(2)})}\\ 
    \times \delta(\Delta - \Delta_{12}) \, ,
\nonumber
\end{align}
where $Dr$ represents the integration over all the configurational degrees of freedom and $Z = \int Dr e^{-\beta H({r})}$ is the partition function.

In presence of ergodicity breaking, the Gibbs measure can be decomposed over \textit{pure states} \cite{mezard1987spin,parisi1988statistical,castellani2005spin}, yielding
\begin{equation}
\label{eq:partition2}
    Z =\sum_{\alpha} \int_{r \in \alpha} Dr e^{-\beta H(r)} = \sum_{\alpha} Z_{\alpha} \, ,
\end{equation}
where
\begin{equation}
    \label{eq:gibbs-split}
    Z_{\alpha} = \int_{r \in \alpha} Dr e^{-\beta H(r)} \, . 
\end{equation}
and $\alpha$ is the pure state index. Whenever Eq.~\ref{eq:partition2} is valid, we can rewrite Eq.~\ref{eq:pdelta} by taking explicitly into account the contribution of different pure states:  
\begin{equation}
    P(\Delta) =  \sum_{\alpha \beta} w_\alpha w_\beta \delta(\Delta - \Delta_{\alpha\beta}) \, ,
    \label{pdelta}
\end{equation}
where $w_\alpha = Z_\alpha/Z$ is the statistical weight of the state $\alpha$. The sum runs over all the minima and includes the case $\alpha=\beta$. We define its average over different realizations of the disorder as $\overline{P(\Delta)} = \overline{ \sum_{\alpha \beta} w_\alpha w_\beta \delta(\Delta - \Delta_{\alpha\beta}) }$.

In the low temperature limit, we can expand the integral in Eq.~\ref{eq:gibbs-split} around the energy $E_{\alpha}$ of the correspondent inherent structure getting, to the first order, $Z_\alpha \sim e^{-\beta E_\alpha}$. Analogously, by repeating the same analysis for the uncompressed region, where the control parameter is the pressure, we get $Z_{\alpha} \sim e^{-Np/\varphi^{\alpha}_J}$ where $p$ is a proxy for the pressure, $N$ is the system size, and $\varphi^{\alpha}_J$ is the jamming packing fraction of state $\alpha$. 

Substituting these approximations into Eq.~\ref{pdelta}, we obtain that the distance distribution is given by 
\begin{equation}
    \begin{cases}
    P_p(\Delta)=\mathcal{N}_p\sum_{\alpha\beta}e^{-Np/\varphi^{\alpha}_J}\,e^{-Np/\varphi^{\beta}_J}\,\delta(\Delta-\Delta_{\alpha\beta})\\
    P_{T}(\Delta) = \mathcal{N}_{T}\sum_{\alpha \beta} e^{-\beta E_\alpha} e^{-\beta E_\beta} \delta(\Delta - \Delta_{\alpha\beta})
    \label{eq-pdelta}
    \end{cases}
\end{equation}
where $\mathcal{N}_p$ and $\mathcal{N}_{T}$ are normalization constants, in both cases equal to the square of the partition function.

Qualitatively, $P(\Delta)$ displays a delta peak at $\Delta=0$, coming from the self-part of the summation ($\alpha=\beta$), and a continuous band that originates from the exchange part ($\alpha\neq\beta$). The former can be used to predict a crossover at which the weight of the entropic term becomes more important than the energetic one. The presence of the continuous band from the exchange part is in agreement with the $d=\infty$ picture which predicts the presence of a huge number of states at jamming due to the Gardner transition. The continuous band contains the details of the organization of the states.

We define the coefficient of the self-part as $W_{T,p}^0 = \sum_\alpha w^2_\alpha$. From Eq.~\ref{eq-pdelta}, we see that $W^0_{T,p}$ becomes more and more important when $T$ decreases or $p$ increases. Indeed, if $T=0$ or $p=\infty$ the only contribution comes from the deepest minimum of the basin with itself and $W^0_{T,p}=1$. On the contrary, at high $T$ or low $p$, we expect to have also the continuous part of the distribution $P(\Delta\neq0)$. When $T=\infty$ or $p=0$ all the minima contribute with the same weight and $W^0_{T,p}=N_{minima}^{-1}$.

In the overcompressed phase, we can easily take into account the harmonic corrections to get
\begin{equation}
    P_{T,vib}(\Delta) = \mathcal{N}_{T,vib} \sum_{\alpha \beta} e^{-\beta  F^{(h)}_\alpha} e^{-\beta F^{(h)}_\beta} \delta(\Delta - \Delta_{\alpha\beta})\, ,
    \label{pdeltacurv} 
\end{equation}
where $F^{(h)}_{\alpha}$ denotes the free energy in the harmonic approximation 
\begin{equation}
    F^{(h)}_{\alpha}= E_{\alpha}+ T\bigl(3\Gamma_{\alpha} - 3N\log(T)\bigr) \, .
    \label{eq-harmonic-free-energy}
\end{equation}

\subsection{Finite Temperature}
\label{sec-pDelta-temperature}

\begin{figure}[h]
    \centering
    \includegraphics[width=\columnwidth]{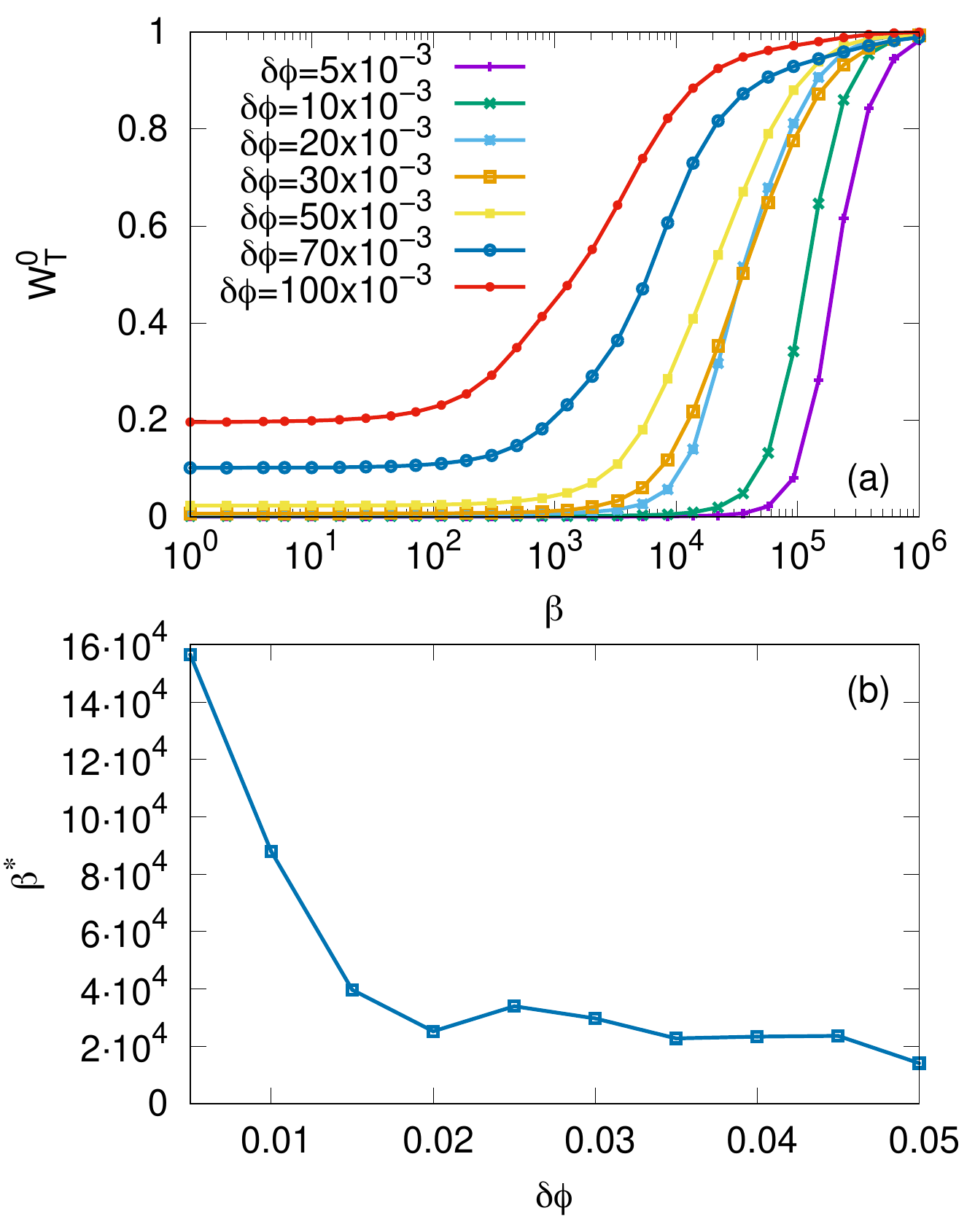}
    \caption{(Color online) (a) $W^0_T$ for several packing fraction values as a function of the inverse temperature $\beta$. (b) Inverse crossover temperature $\beta^*=(T^*)^{-1}$ as a function of the excess packing fraction $\delta\varphi$.}
    \label{fig:betaeff}
\end{figure}

\noindent In Fig.\hyperref[fig:betaeff]{~\ref{fig:betaeff}(a)}, we show the behavior of $W^0_T$, the coefficient of the self-part, as a function of the temperature. We observe the presence of a crossover between the low and the high temperature behavior, where we go from a situation in which the thermodynamic properties are determined by a finite number of states, whose energy is close to that of the deepest minimum, to a regime in which all the minima are equally important.

We define the crossover temperature $T^*$ via the condition $W^0_{T^*}=1/3$. The result is shown in Fig.\hyperref[fig:betaeff]{~\ref{fig:betaeff}(b)}: we see that, as the packing fraction decreases, and consequently the number of states increases, smaller temperatures are needed to overcome the entropic contribution. From Fig.\hyperref[fig:betaeff]{~\ref{fig:betaeff}(a)} we also see that, decreasing $\delta\varphi$, the crossover becomes steeper. 

We notice that, since this crossover involves just a finite number of states of low energy, it is not affected by the plateau at high energies found in the cumulative distributions of the number of inherent structures (Fig.~\ref{fig:complexity}); in other words, the presence in the cage of a number of inherent structures larger than what can be sampled by our algorithm, affects only Fig.\hyperref[fig:betaeff]{~\ref{fig:betaeff}(a)} at very small values of $\beta$, but leaves the qualitative picture and the determination of $\beta^* = 1/T^*$ intact.

\begin{figure}[h]
    \centering
    \includegraphics[width=\columnwidth]{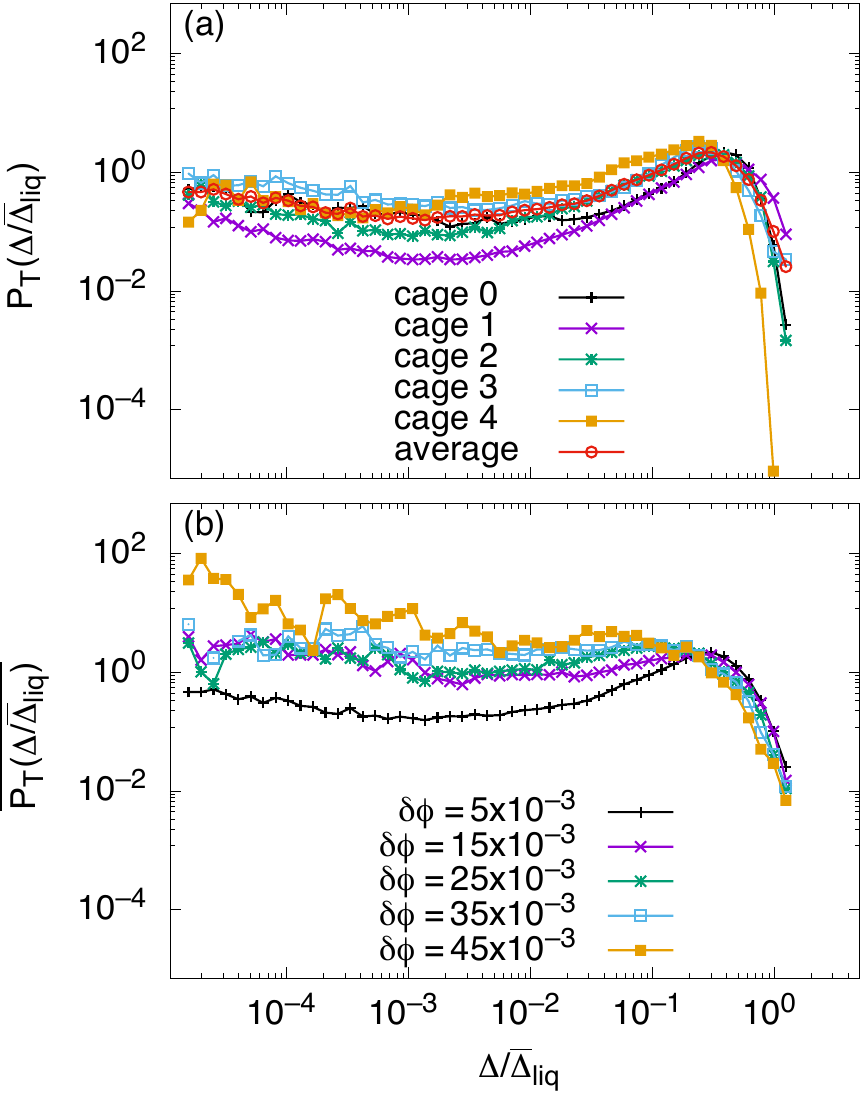}
    \caption{(Color online) (a,b) Distance distributions at $\beta=0$ as a function of $\Delta/\overline{\Delta_{\text{liq}}}$, where $\overline{\Delta_{\text{liq}}}=7.3\times10^{-3}$. In particular, (a) $P_T(\Delta/\overline{\Delta}_{\text{liq}})$ at $\delta\varphi=5 \times 10^{-3}$ has large sample-to-sample fluctuations that, however, do not affect the qualitative shape of the distribution. (b) $\overline{P_T(\Delta/\overline{\Delta}_{\text{liq}})}$ at various $\delta\varphi$. Upon decreasing the packing fraction, states at small distances are observed more rarely because geometrically disadvantaged.
    Since the singular term at $\Delta=0$ has been taken into account, all the probability distributions are normalized to $1-W^0_T$.}
    \label{fig:pdelta1}
\end{figure}

Now, let us focus on the continuous branch of the displacement distribution $P_T(\Delta)$. In Fig.\hyperref[fig:pdelta1]{~\ref{fig:pdelta1}(a)} we show the distributions at $\beta=0$ and $\delta \varphi = 5 \times 10^{-3}$ for all the samples. We find that the distribution has support over a continuous range of $\Delta$ values, covering nearly 5 orders of magnitude, compatibly with the existence of many states as predicted by the mean-field theory. $P_T(\Delta)$ exhibits a nearly flat behavior. We notice that, as already highlighted in Fig.~\ref{fig:complexity_fluct}, there are large sample-to-sample fluctuations, but the overall shape of the probability distribution is preserved.

\begin{figure}[h]
    \centering
    \includegraphics{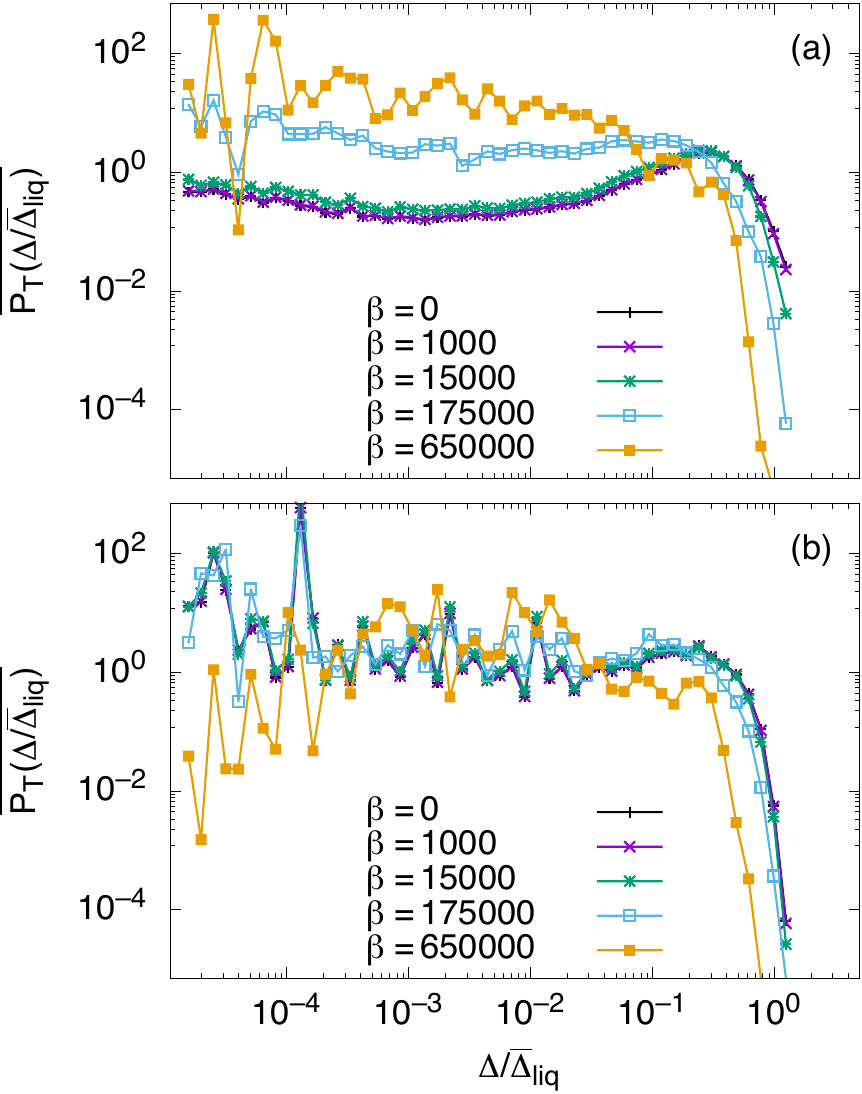}
    \caption{(Color online) Thermal behaviour of the average distance distribution $\overline{P_T(\Delta/\overline{\Delta}_{\text{liq}})}$, where $\overline{\Delta_{\text{liq}}}=7.3\times10^{-3}$. (a) Distance distribution averaged over all the samples at  $\delta\varphi = 5 \times 10^{-3}$ for several $\beta$ values: cooling enhances the probability of finding states at small $\Delta$. (b) Same plots of (a), including the harmonic corrections of Eqs.~\ref{pdeltacurv}-\ref{eq-harmonic-free-energy}. Their effect is to make more uniform the behavior of the distribution at different temperatures. Distributions are normalized to $1- W^0_T$.}
    \label{fig:pdelta2}
\end{figure}

In Figs.~\hyperref[fig:pdelta1]{~\ref{fig:pdelta1}(b)},\hyperref[fig:pdelta2]{~\ref{fig:pdelta2}(a)} we consider how the distribution $\overline{P_T(\Delta)}$ modifies when the packing fraction or the temperature is changed. Here, $\overline{\bullet}$ indicates the average over all the samples. We see that the probability of finding two local minima of the landscape at smaller $\Delta$ increases when the temperature is lowered or the packing fraction is increased. The former can be easily explained by observing that the Boltzmann weight $e^{-\beta E}$ strongly suppresses the pairs of configurations with the highest energy, that in our system also correspond to those at a larger distance. The dependence on the packing fraction can be understood by noticing that, given any minimum, the number of neighboring minima increases with the distance. Since, decreasing $\delta\varphi$, the landscape roughens and the number of local minima increases, there will be more configurations at large distances giving rise to a peak at high $\Delta$ values.

We also notice that in Fig. \hyperref[fig:pdelta2]{\ref{fig:pdelta2}(a)} the probability distribution is rather insensible to temperature changes up to the crossover temperature, where the states close to the ground state start to dominate the partition function.

Finally, in Fig.~\hyperref[fig:pdelta2]{\ref{fig:pdelta2}(b)} we present $\overline{P_{T,vib}(\Delta)}$ for several temperatures, where we have included the harmonic terms as prescribed by Eqs.~\ref{pdeltacurv}-\ref{eq-harmonic-free-energy}. Their effect is to make more uniform the behavior of the distribution at different $\beta$'s, by enhancing small $\Delta$ probabilities. However, the qualitative picture of $\overline{P_T(\Delta)}$ is not altered.

\subsection{Finite Pressure}

\noindent In analogy with Secs.~\ref{subsec-jamming-structure}-\ref{subsec-hierarchical-structure}, in the uncompressed region we compute $P_p(\Delta)$ considering only the 1000 deepest minima of each sample.

\begin{figure}[htp]
\centering
\includegraphics[width=\columnwidth]{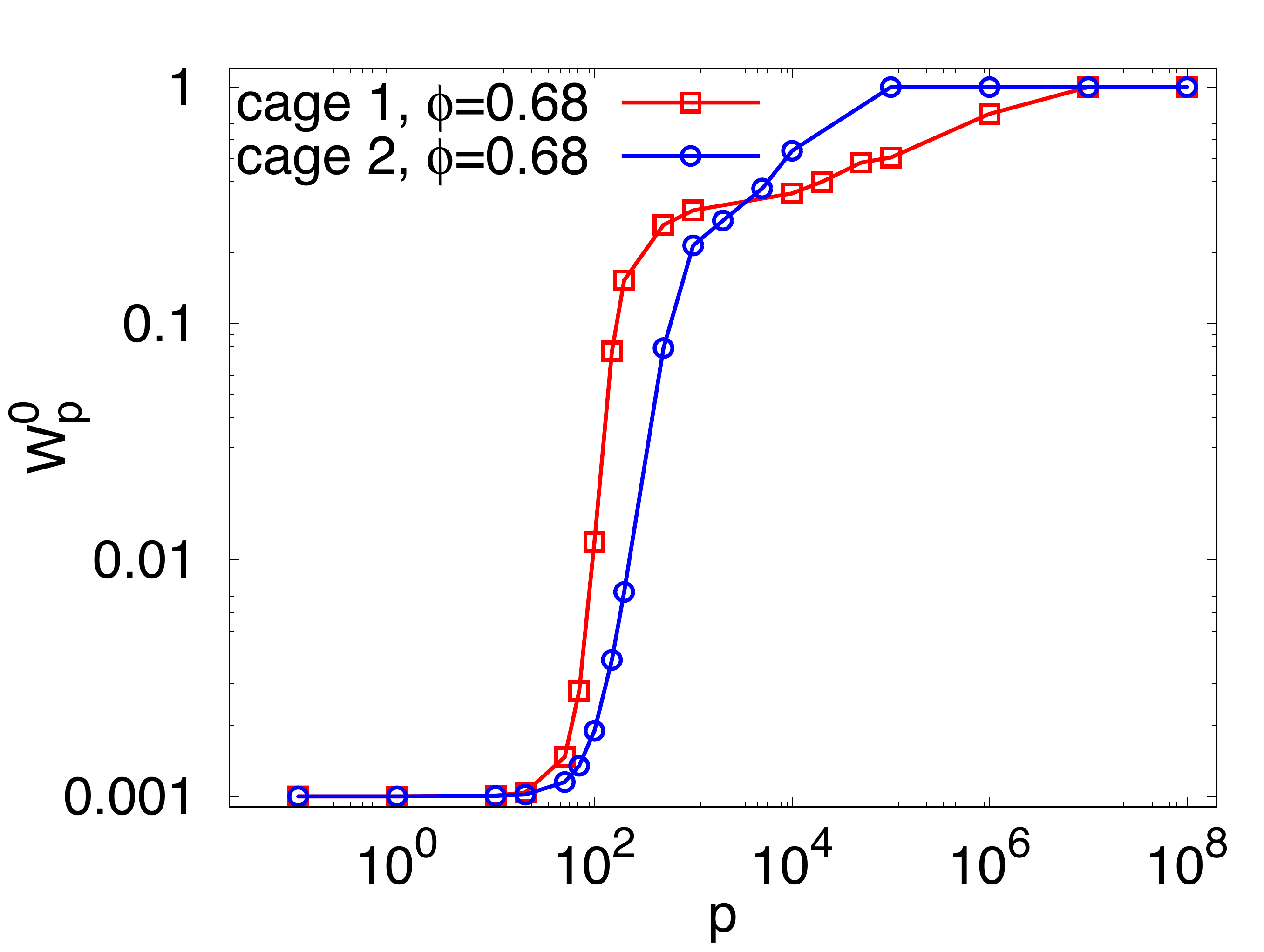}
\caption{(Color online) $W^0_p$ for: (a) cage 1 generating the clones at $\varphi=0.68$; (b) cage 2 generating the clones at $\varphi=0.68$. We consider only the 1000 deepest minima of the samples. $W^0_p=10^{-3}$ at low pressure because all the jamming local minima have the same weight; while it saturates to 1 when the only contribution to $P_p(\Delta)$ comes from the deepest minimum of the basin.}
\label{fig-p0-jamming-68}
\end{figure}

In Fig.~\ref{fig-p0-jamming-68} we present the behavior of $W^0_p$ as a function of the pressure for two samples reaching the jamming point from $\varphi=0.68$. As in Sec.~\ref{sec-pDelta-temperature}, this plot allows us to determine the crossover from the entropy ruled and the ground state ruled landscape.

\begin{figure}[htp]
\centering
\subfloat[]
{\includegraphics[width=\columnwidth]{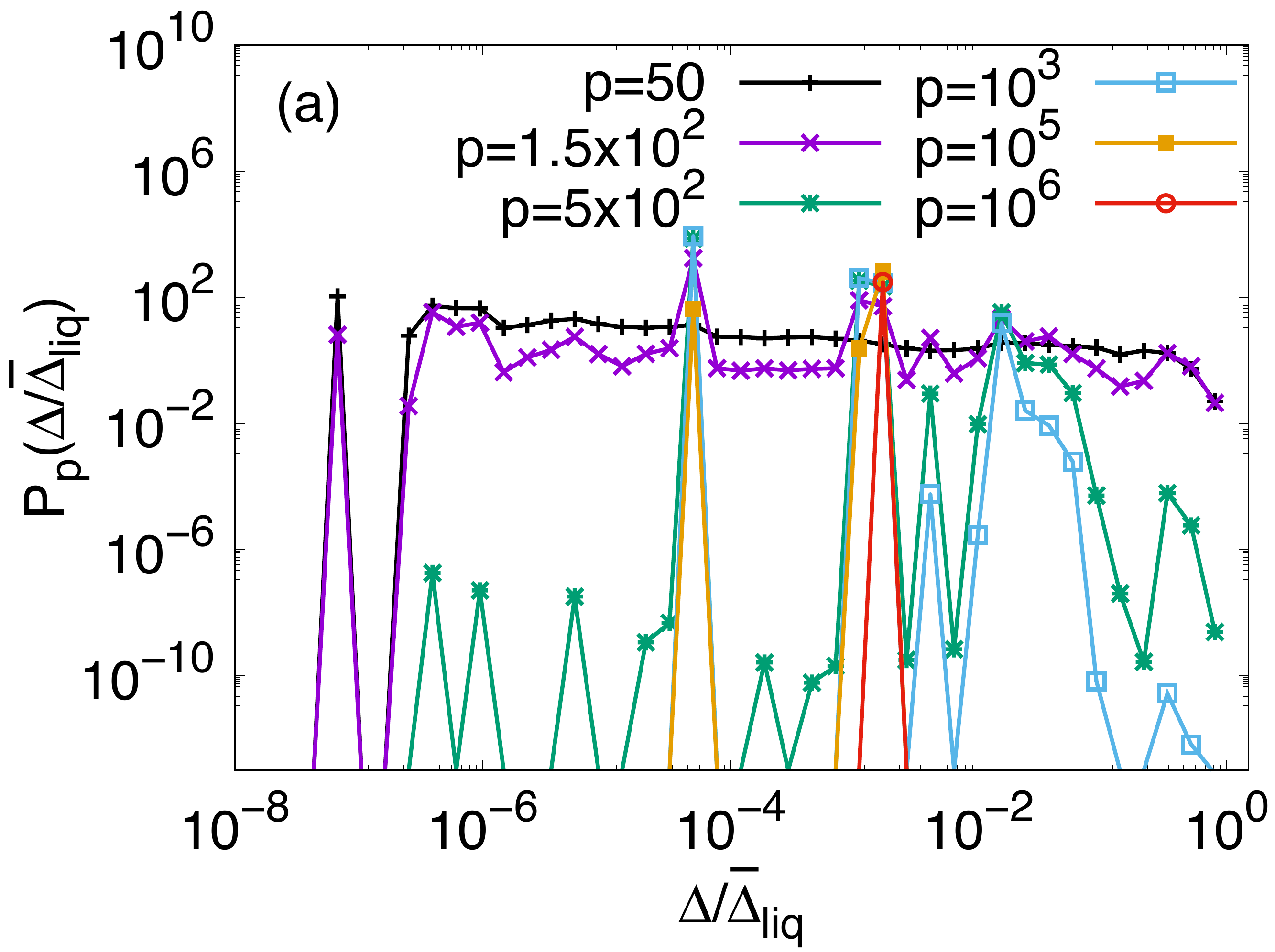}
\label{fig-pdelta-jamming-conf1-68}}
\vspace{-6mm}
\subfloat[]
{\includegraphics[width=\columnwidth]{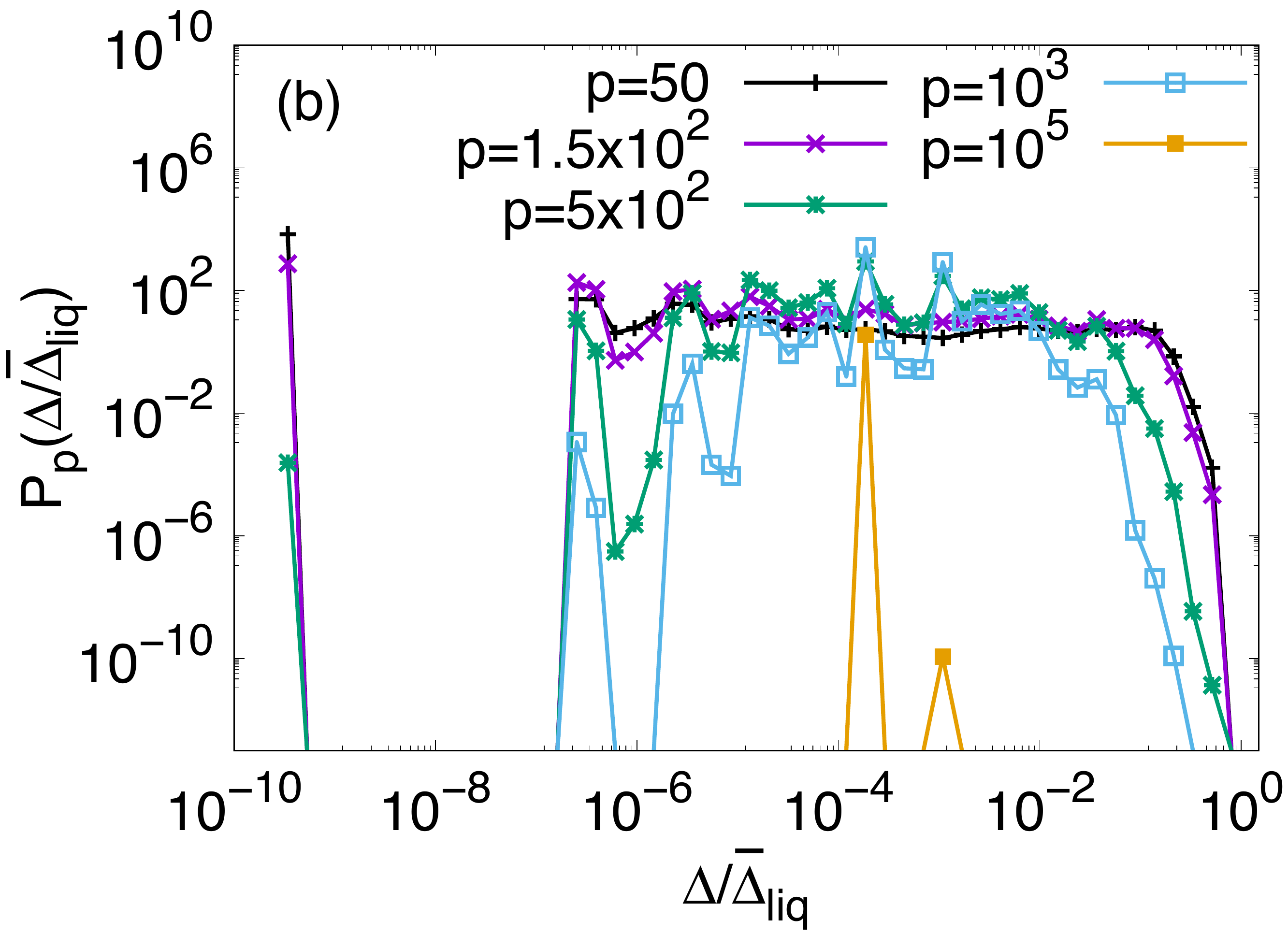}
\label{fig-pdelta-jamming-conf2-68}}
\caption{(Color online) $P_p(\Delta/\overline{\Delta_{\text{liq}}})$ for the same samples studied in Fig.~\ref{fig-p0-jamming-68} constructed with the 1000 deepest minima of the samples. On the $x$ axis, $\Delta$ divided by $\overline{\Delta_{\text{liq}}}=7.3\times10^{-3}$. The distributions are normalized such that ${\int d\Delta\, P_p(\Delta)=1-W^0_p}$. (a) Results for cage 1 generating the clones at $\varphi=0.68$. (a) Results for cage 2 generating the clones at $\varphi=0.68$.}
\label{fig-pdelta-jamming-68}
\end{figure}

In Fig.~\ref{fig-pdelta-jamming-68} is shown the $P_p(\Delta)$ for the same samples of Fig.~\ref{fig-p0-jamming-68}. In both cages, the probability $P_p(\Delta)$ at low $p$ is almost flat. Furthermore, as expected, the distribution presents isolated peaks when $p$ increases.

How the distribution changes with pressure is different in each cage because of the high heterogeneity of the landscape basins. For instance, in cage 1 at $p=10^6$ $P(\Delta\neq0)\neq0$ because the sum in Eq.~\ref{eq-pdelta} has a residual contribution from the second deepest minimum, which has $\varphi_J$ very close to the one of the ground state. At low pressure, all the minima contribute.
\section{\label{sec-conclusions} Conclusions and Outlook}

\noindent This exploratory study aims to investigate the structure of the landscape near the jamming transition in three-dimensional soft spheres systems, in the light of the existence of the Gardner transition in the exact solution of hard spheres in $d=\infty$. We followed two different but complementary approaches. 

In the first one, we brought the system to the jamming point from the uncompressed phase, starting from two different packing fraction values. This approach pointed out the presence of subcages and of a huge number of local minima at jamming, in agreement with the fractal nature of the landscape in $d=\infty$. We found that, in many cases, the deepest minima are close in the landscape and located in a deep well. They usually have large basins of attraction. Instead, the minima corresponding to smaller values of $\varphi_J$ are scattered in all the jamming landscape and they usually have small basins of attraction. The ultrametric structure of the landscape and the clustering properties of the jammed configurations seem to be verified by the high value of the AC coefficients, even though tested on a small bunch of minima. Furthermore, our study shows the possibility of using in an efficient way a Linear Programming (LP) algorithm to reach the jamming point in finite-dimensional hard spheres system.

On approaching the jamming transition from the overcompressed region, we found a new kind of criticality: the number of inherent structures as a function of the packing fraction seems to diverge in a power-law fashion when $\varphi\rightarrow\varphi_{J}$, and the cumulative of the number of inherent structures at low energies behaves as a power-law in a wide range of packing fraction values, with a $\varphi$-dependent exponent. Moreover, the onset of the critical behavior is accompanied by the emergence of large sample-to-sample fluctuations, which can be seen both in the cumulative of the inherent structures and in the distribution of the curvatures from the Hessian matrix. Furthermore, the latter displays a Gaussian behavior.

The study of the distribution of the distances both in the uncompressed and overcompressed region, as a function of $p$ and $T$ respectively, revealed the existence of a crossover between ground state ruled and entropy ruled landscapes. We found the presence of an isolated peak at zero distance and of a continuous part of the distribution at larger distances. The latter displays an almost flat behavior across several values of pressure, temperature and packing fraction. The existence of a continuous part extended over five orders of magnitude shows that the landscape is made up of a huge number of states which have a continuous distribution of distances, in agreement with the $d=\infty$ picture. The inclusion of harmonic corrections does not change in a significant way the behavior of the distribution that is, therefore, correctly captured already at a purely energetical level.

The pictures of the landscape that emerge from the study at jamming and in the overcompressed region are in agreement and consistent with the mean-field predictions. However, we observe that quantities like $P(\Delta)$ are very atypical and, therefore, a conclusive analysis should be done over a larger number of local minima and averaging over several different cages. Indeed, increasing the number of starting metabasins, would allow to perform better disorder averages and have deeper control over sample-to-sample fluctuations. 

Moreover, repeating the same analysis with systems of different sizes and dimensions would allow us to determine the entity of finite-size effects and to properly extrapolate the thermodynamic limit of the observables.
 
Furthermore, since it is known \cite{berthier2016growing,charbonneau2015numerical} that the starting point of the compression plays an important role in determining the strength of the Gardner transition, it would also be important to repeat the analysis by starting from different points along the equation of state and identify, in a clear way, the universal features of the  behavior at jamming

\section{Acknowledgements}
\noindent We thank L. Berthier for providing the initial configurations and useful exchanges. We thank P. Charbonneau, G. Folena, C. Scalliet, A. Scardicchio, G. Sicuro and G. Tsekenis for many useful discussions related to this work. The research has been supported by the Simons Foundation (grant No. 454949, G. Parisi).

\appendix
\section{\label{app-lp} The Linear Programming Algorithm}

\noindent A Linear Programming algorithm is a method to solve linear optimization problems. An LP problem is the problem of \textit{minimization or maximization of a linear function subjected to linear constraints} \cite{dantzig2006linear,dantzig2006linear2}. 
Our LP problem has the objective to maximize the particles' radii. The inflation rate has to be equal for all the particles to leave unchanged the degree of polydispersity. The variables are the inflation rate $\alpha$ and the particles' displacements $\boldsymbol{\Delta}$ ($3(N-1)$ scalar variables) \footnote{We fix the position of a sphere to avoid the rigid translation of all the system.}. The constraints of the problem are

\begin{align}
\label{eq-constraint-lp}
    (\boldsymbol{x}_i-\boldsymbol{x}_j)^2+2(\boldsymbol{x}_i-\boldsymbol{x}_j)(\boldsymbol{\Delta}_i-\boldsymbol{\Delta}_j)+\\
    \notag
    +(\boldsymbol{\Delta}_i-\boldsymbol{\Delta}_j)^2-\alpha\sigma^2_{ij}\leq 0
\end{align}
where $\sigma_{ij}=\frac{\sigma_i+\sigma_j}{2}$ is the sum of the radii, $\mid \Delta^{x,y,z}_i \mid\leq c$ and $0\leq\alpha\leq c'$ for $c$ and $c'$ reasonable values with respect to the linear dimensions of the system. 

To use LP we need a linear problem. We can neglect the term $(\boldsymbol{\Delta}_i-\boldsymbol{\Delta}_j)^2$ supposing that the displacements are small. This assumption is asymptotically justified because, after some iterations, the increase of the particles' radii will reduce the magnitude of the possible displacements. In the end, the problem can be written as a maximization problem with objective function $\alpha$.

\begin{figure}[htp]
\centering
\includegraphics[scale=0.15]{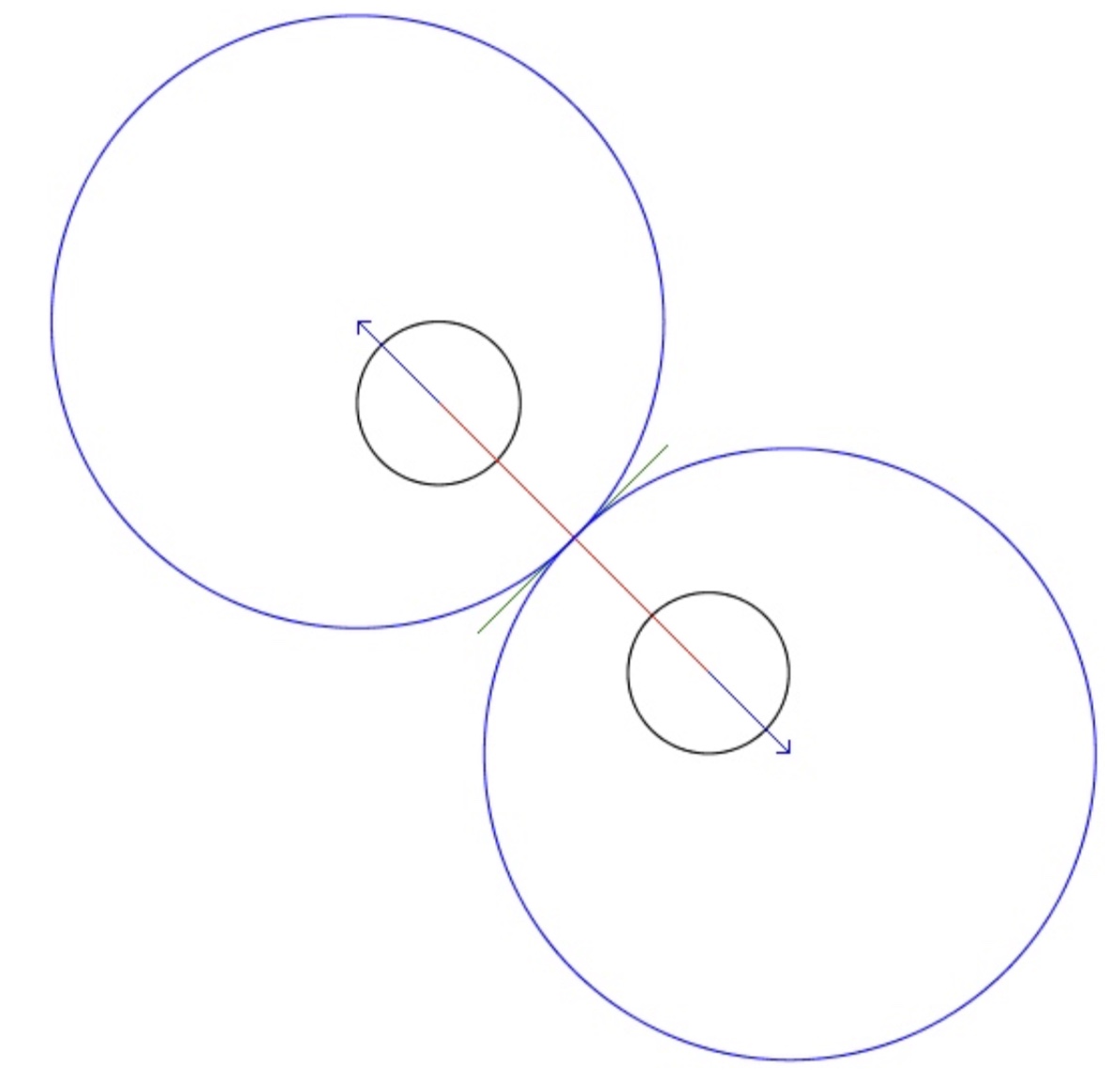}
\caption{(Color online) Planar representation of the LP optimal solution for the compression (\textit{i.e.} inflation) of two hard spheres. In red, the segment joining the two spheres, in green the LP constraint and in blue the arrow of the optimal displacement.}
\label{fig-constraint-lp}
\end{figure}

For each couple of particles, the constraint in Eq.~\ref{eq-constraint-lp} is satisfied and the inflation $\alpha$ is maximized when the spheres go as far as possible along the direction orthogonal to the constraint and they are inflated until the constraint is saturated (Fig.~\ref{fig-constraint-lp}). Since each sphere has many neighbors and all the constraints have to be satisfied simultaneously, the displacements would not be only along this orthogonal direction and the constraint will not be saturated after only one iteration. We need some iterations ($\sim\mathcal{O}(10)$) to reach the jamming point. 

The LP algorithm we used is in the \texttt{GLPK} library, version 4.55 \footnote{\texttt{GLPK} uses the symplectic method. The Reference Manual can be found here \cite{glpk}.}.

\section{\label{app-rattler} Rattlers}

\noindent In jammed hard spheres packings in finite dimensions some particles are not part of the force network. They are called \textit{rattlers} and they have $Z<d+1$ contacts. Typically, their neighbors are part of the force network and form a cage in which the rattlers can freely move. In this study, we compute the observables excluding the rattlers. The distance between two packings is computed as the distance of their backbones intersection (\textit{i.e.} we excluded the rattlers in common and not in common between the two packings).

In a polydisperse system, the number of rattlers is bigger than in a monodisperse one. Indeed, it is easier for a particle with a small radius to be trapped by the others. At jamming, we find that the average number of rattlers per jammed configuration is $\overline{N}_{rattlers}=15$.

With the LP algorithm of the \texttt{GLPK} library, version 4.55, after each iteration, one finds that most of the rattlers computed on the LP constraints have $3$ contacts. This means that they are leaned against the walls of the virtual box of the constraints of the LP problem. This is not what one would expect because rattlers should be free to move and, so, they typically have 0 contacts. Furthermore, the final jammed packings are not isostatic on the physical contacts. We believe this result is due to a computer rounding on the forces acting on the rattlers, which moves them arbitrary.  

To solve this problem, we define a smooth potential which, minimized at each LP iteration, moves the rattlers in the middle of their virtual box. This potential is defined for each group of nearest rattlers. In the case of an isolated rattler, the potential is defined as

\[
F(\boldsymbol{h},\epsilon)=-\sum^{N'}_{i=1}
\begin{cases}
\ln{(h_i)}       & \text{if $h_i>\epsilon$} \\
[\frac{h_i-\epsilon}{\epsilon}+\ln{(\epsilon)}] & \text{if $h_i<\epsilon$}
\end{cases}
\]
where the sum is over all the $N'$ non-rattlers of the packing, $h_i$ is the gap between the considered rattler and the particle $i$ and $\epsilon=10^{-12}$. The generalization to the case of a group of rattlers is straightforward. For the minimization of the potential, we use the \texttt{nmsimplex2} algorithm in the \texttt{GSL} library \cite{gsl}.

In this way, the resulting jammed configurations are isostatic in the physical gaps, except for a small percentage of non-isostatic packings \footnote{The percentage of non-isostatic packings is $9\%$. The percentage is much smaller excluding from the average the second cage (both starting from $\varphi=0.647$ and $\varphi=0.68$) and becomes $2\%$. The jammed configurations of cage 2 are more difficult to reach with our LP algorithm. This might be due to the presence of more rattlers than in the other cages.}.
Although the potential expedient, we find that the rattlers' positions in the jammed packings are random. Indeed, looking at the rattlers in different clones that have reached the same local minimum, they have different positions. Hence, to avoid spurious contributions from the rattlers we compute the distances between two packings as the distance of the backbones intersection (see App.~\ref{app-distances}).

\section{\label{app-distances} Distance among Jammed Packings}

\noindent The distance of two configurations is defined as (Eq.~\ref{eq-square-distance})
\begin{equation}
\label{eq-delta-app}
    \Delta=\frac{1}{M}\sum^{M}_{i=1}(\boldsymbol{x}^{\alpha}_i-\boldsymbol{x}^{\beta}_i-\boldsymbol{\delta})^2
\end{equation}
while the overlap is (Eq.~\ref{eq-overlap})
\begin{equation}
\label{eq-Q-app}
    Q=\frac{1}{M}\sum_{i,j}^{1,M}\Theta\biggl(a-\mid\boldsymbol{x}_i^{\alpha}-\boldsymbol{x}_j^{\beta}-\boldsymbol{\delta}\mid\biggr)
\end{equation}
$\alpha$ and $\beta$ are the minima indeces, $i$ is the particle index, $M$ is the number of spheres in the backbones intersection, \textit{i.e.} $M$ is the total number of particles which are non-rattlers neither in packing $\alpha$ nor in $\beta$, and $\boldsymbol{\delta}$ is the distance of the centers of mass of $\alpha$ and $\beta$. 

Indeed, as highlighted in Sec.~\ref{sec-local-minima}, when we compute observables such as $\Delta$ and $Q$ we need to avoid spurious effects coming from the system translational invariance. Hence, to compute $\Delta$ and $Q$ we perform a rigid translation in order to have $\alpha$ and $\beta$ with coincident centers of mass: we compute the average displacement of the particles among the two configurations as $\boldsymbol{\delta}=\frac{1}{M}\sum_{i=1}^{M}(\boldsymbol{x}_i^{\alpha}-\boldsymbol{x}_i^{\beta})$ and we subtract $\boldsymbol{\delta}$ in each term of the sum $\boldsymbol{x}^{\alpha}_i-\boldsymbol{x}^{\beta}_i$. From Eqs.~\ref{eq-delta-app},~\ref{eq-Q-app} it is clear that, in this way, packings representing the same local minima have $\Delta=0$ and $Q=1$.

In the LP protocols, before doing any analysis, we computed the distance distribution among the 1000 deepest minima in each sample. Studying them in log scale, we found that at very small distances the distributions present some isolated peaks. Moreover, the same feature is present in the log distribution of the packing fraction differences $\Delta\varphi_J$. After a careful analysis, we argued that configurations with both a very small $\Delta$ and a very small $\Delta \varphi_J$ should be considered  \textit{degenerate} because they represent the same local minimum of the landscape. Looking at those log distributions, we fixed two threshold values $\Delta^{*}\leq10^{-9}$ and $\Delta \varphi_J^*\leq10^{-9}$. Whenever two configurations have $\Delta\leq\Delta^{*}$ and $\Delta \varphi_J\leq\Delta \varphi_J^*$, we considered them as degenerate.

\bibliography{bibliography.bib}

\end{document}